\documentclass[preprint,showpacs,preprintnumbers,amsmath,amssymb]{revtex4}

\usepackage{graphicx}
\usepackage{dcolumn}
\usepackage{bm}

\begin{document}


\title{Light nuclei quasiparticle energy shifts in hot and dense nuclear matter}

\author{ G. R\"opke}
\affiliation{
Universit\"at Rostock, Institut f\"ur Physik, 18051 Rostock, Germany}

\date{\today}

\begin{abstract}
Nuclei in dense matter are influenced by the medium. In the cluster mean field approximation, an  effective Schr\"odinger equation for the $A$-particle cluster is obtained accounting for the effects of the correlated medium such as
self-energy,  Pauli blocking and Bose enhancement. Similar to the single-baryon states (free neutrons and protons), the light elements ($2 \le A \le 4$, internal quantum state $\nu$) are treated as quasiparticles with energies  $E_{A,\nu}(\vec P; T, n_n,n_p)$. These energies depend on the center of mass momentum $\vec P$, as well as temperature $T$ and the total densities $n_n,n_p$ of neutrons and protons, respectively. No $\beta$ equilibrium is considered so that $n_n, n_p$ (or the corresponding chemical potentials $\mu_n, \mu_p$) are fixed independently. 

For the single nucleon quasiparticle energy shift, different approximate expressions such as Skyrme or relativistic mean field approaches are well known.  Treating the $A$-particle problem in appropriate approximations, results for the cluster quasiparticle shifts are given. Properties of dense nuclear matter at moderate temperatures in the subsaturation density region considered here are  influenced by the composition. This in turn is determined by the cluster quasiparticle energies, in particular the formation of clusters at low densities when the temperature decreases, and their dissolution due to Pauli blocking as the density increases. Our finite-temperature Green function approach covers different limiting cases: The low-density region where the model of nuclear statistical equilibrium and virial expansions can be applied, and the saturation density region where a mean field approach is possible. 
\end{abstract}

\maketitle

 \section{\label{sec:introduction}
 Introduction}

A well established tool in treating many-particle systems is the quasiparticle concept. In contrast to free particles, the  properties of quasiparticles such as the dispersion relation are modified due to the interaction with other particles. This relation between energy and momentum is often characterized by an energy shift and an effective mass which depends on  temperature and density of the medium. A significant part of the interaction can be taken into account by introducing this approach. The quasiparticle concept is not restricted to elementary particles like nucleons only, but it can also be applied to composed particles, i.e. nuclei. (Note that the nucleons itselves are also composed particles.)  In this work we evaluate these cluster quasiparticle energies within a microscopic approach.

Nuclear matter is a strongly interacting quantum fluid. In order to treat warm and dilute matter (i.e. at subsaturation baryon densities $n_B =n_n+n_p \lesssim 0.16 \,\,{\rm fm}^{-3}$ and temperatures $T\lesssim 20$ MeV) 
within a systematic quantum statistical approach,   we start from a nonrelativistic  Hamiltonian
\begin{equation}
\label{Ham}
 H = \sum_1 E(1)a_1^\dagger a_1^{} +\frac{1}{2} \sum_{12,1'2'} V(12,1'2') a_1^\dagger a_2^\dagger a_{2'}^{} a_{1'}^{}\,,
\end{equation}
where \{1\} denotes momentum $\hbar \, \vec P_1$, spin $\sigma_1$, and isospin $\tau_1$ characterizing the neutron ($n$) or proton ($p$) state. The kinetic energy in fermion second quantization $a_1^\dagger, a_1^{}$ contains $E(1)=\hbar^2 P_1^2 / 2 m_1$, whereas the potential energy contains the matrix element $V(12,1'2') $ of the nucleon-nucleon interaction.

Since there is no fundamental expression for the nucleon-nucleon interaction (like, e.g., the Coulomb interaction in charged particle systems \cite{KKER}), it is taken to reproduce empirical data such as the nucleon scattering phase shifts. Different  parametrization are in use. Simple potentials as proposed by Yukawa, Yamaguchi, Mongan, Gogny and others are based on two-nucleon phase shifts and can be used for exploratory calculations. For detailed calculations one can use more sophisticated potentials such as PARIS and BONN or their separable representations \cite{PEST}. To obtain the empirical parameter values of nuclear matter at saturation density, three-body forces have been introduced in the Hamiltonian (\ref{Ham}). In particular, the Argonne AV18/UIX potential \cite{Wir95} has been used to calculate light nuclei \cite{Wir01}.

Near saturation density, nucleons can be treated as quasiparticles. Semiempirical approaches such as the Skyrme contact pseudopotential \cite{Vau72} and relativistic mean field (RMF) approaches \cite{Typel} parametrize the quasiparticle shift as a function of densities and temperature. Microscopic approaches such as Dirac-Brueckner Hartree Fock (DBHF) \cite{fuc06} give an appropriate description of the thermodynamic properties of warm and dense matter. For a recent review of the nuclear matter equation of state (EoS) see Ref. \cite{Klahn:2006ir}.

In the low-density limit, nuclear matter at finite temperature is a mixture of free nucleons and nuclei in chemical equilibrium as described by a mass action law. This chemical picture, where bound states are treated as new species, is also denoted as nuclear statistical equilibrium (NSE) and should be recovered as the low-density limit of  a quantum statistical approach to nuclear matter at finite temperatures.  

At increasing densities, the simple NSE approach becomes invalid because of interactions between the nucleons and nuclei. A systematic coherent description should include scattering states, avoid double counting of many-particle effects, and avoid semiempirical concepts such as excluded volume. This can be achieved within a physical picture where bound states are produced by the interaction, investigating the few-particle propagator in the medium. The chemical picture can be used as a guideline to select important contributions within a perturbative approach and performing partial summations of the corresponding Feynman diagrams. In the case of charged particle systems, where the interaction is given by the Coulomb potential, the use of the chemical picture to introduce bound states in the partially ionized plasma has been extensively worked out, see Ref.  \cite{KKER}.

The in-medium wave equation for a system of $A$ nucleons is derived in Sec. \ref{sec.2}. Bound states describe nuclei  with energy eigenvalues $E_{A,\nu}(P;T,n_p,n_n)$ where $A$ denotes the mass number.  The index $\nu$ specifies the internal state of the $A$-nucleon system such as spin, isospin and excitation. In a homogeneous system, the center-of-mass momentum $\vec P$ of the cluster is conserved and can be used as quantum number. Assuming (local) thermal equilibrium, due to the influence of the surrounding matter these energy eigenvalues will depend on three parameter:  The temperature $T$ and the total number densities $n_p, n_n$ of protons and neutrons, respectively [or the baryon density $n_B=n_n+n_p$ and the asymmetry parameter $\alpha=(n_n-n_p)/n_B$].  We do not consider $\beta$-equilibrium due to weak interaction processes. Explicit results for the corresponding quasiparticle shifts are given for $^2$H (deuteron $d$),  $^3$H (triton $t$),  $^3$He (helion $h$), and $^4$He ($\alpha$-particle) in Sec. \ref{quasiparticle shifts}. As an application, the generalized Beth-Uhlenbeck formula is outlined in Sec. \ref{sec:BU}, and the nuclear matter EoS is considered which contains the quasiparticle energies of the single-nucleon states as well as of the light nuclei.

\section{Many particle approach}
\label{sec.2}

\subsection{Single-particle spectral function and quasiparticles} 
\label{sec2.1}

The basic equations of a Green function approach to
the many-nucleon system can be found in different textbooks and papers, see \cite{FW,SRS}. Here we give only some final results.  

The propagation of a single-nucleon excitation $\langle a^\dagger_1(t) a^{}_{1'} \rangle$ can be expressed in terms of the single-particle spectral function $S_1(1,\omega)$ as
\begin{equation}
\langle a^\dagger_1(t) a^{}_{1'} \rangle =\delta_{1,1'} \int \frac{d \omega}{2 \pi} e^{i \omega t} f_{1,Z} (\omega) S_1(1,\omega)\,,
\end{equation}
where  
\begin{equation}
\label{faz}
f_{A,Z}(\omega) =
[\exp( \beta (\omega -Z \mu_p -(A-Z) \mu_n) )-(-1)^A]^{-1}
\end{equation}
is the Fermi or Bose distribution function which depends on the inverse temperature $\beta = 1/(k_BT)$ and the chemical
potentials $ \mu_p,  \mu_n$ (instead of the isospin quantum number $\tau_1$ we use the charge number $Z$).
A special case is the single-particle density matrix $\langle a^\dagger_1 a^{}_{1'} \rangle$ which can be used to evaluate the equation of state (EoS) for the nucleon density 
\begin{equation}
n_{\tau}(\beta,\mu_p,\mu_n)=\frac{1 }{ \Omega} \sum_{1} \langle a^\dagger_1 a^{}_{1} \rangle \delta_{\tau,\tau_1} = 2  \int \frac{d^3P_1}{(2 \pi)^3}
\int_{-\infty}^\infty\frac{d \omega}{2 \pi} f_{1,Z}(\omega) S_1(1,\omega)\,, 
\label{eosspec}
\end{equation}
where $\Omega$ is the system volume, and summation over spin direction
is collected in the factor 2. Both the Fermi distribution function and the spectral function depend on the temperature and the chemical potentials $\mu_p, \mu_n$ not given explicitly. We work with a grand canonical ensemble and have to use the EoS (\ref{eosspec}) to replace the chemical potentials by the densities $n_p,n_n$. For this equation of state, expressions such as the Beth-Uhlenbeck formula are discussed below in Sec. \ref{sec:BU}.

The spectral function is related to the self-energy according to
\begin{equation}
\label{spectral}
 S_1(1,\omega) = \frac{2 {\rm Im}\Sigma(1,\omega-i0)} 
{[\omega - E(1)- {\rm Re} \Sigma(1,\omega)]^2 + 
[{\rm Im}\Sigma(1,\omega-i0)]^2 }\,,
\end{equation}
where the imaginary part has to be taken for a small negative
imaginary part in the frequency.
The solution of the relation 
\begin{equation}
E_1^{\rm qu}(1) = E(1) + {\rm Re} \Sigma[1,E_1^{\rm qu}(1)] 
\end{equation}
defines the single-nucleon quasiparticle energies $E_1^{\rm qu}(1) = E(1) + \Delta E^{\rm SE}(1)$. Expanding for small Im$\Sigma(1,z)$, the spectral function
yields a $\delta$-like contribution, and the densities are calculated from Fermi
distributions with the quasiparticle energies so that
\begin{equation}
n^{\rm qu}_{\tau}(\beta,\mu_p,\mu_n)=\frac{2 }{ \Omega} \sum_{P_1} f_{1,Z}[E_1^{\rm qu}(1)]\,
\label{nqu}
\end{equation}
follows for the EoS in mean field approximation.
This result does not contain the contribution of bound states and therefore fails to be correct in the low-temperature, low-density limit where the NSE describes the nuclear matter EoS.

\begin{figure}
\includegraphics[width=0.8\textwidth]{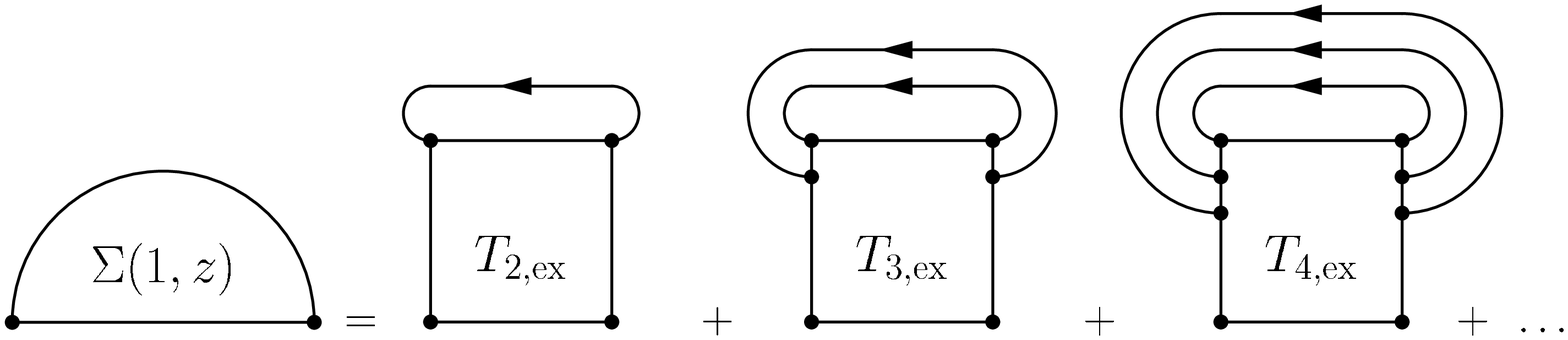}
\caption{\label{fig:clusterdecomp} Cluster decomposition of the single-nucleon self-energy. The index 'ex' denotes full antisymmetrization including all exchange diagrams.
For bound states with $A \le 4$ the direct ladder-T matrices can be taken, without the full exchange at the end, as long as the particles have different internal quantum numbers $\sigma, \tau$. The Fock term results from T$_{2, \rm ex}$ (first order in $V$, exchange term, see also App. \ref{app.1}).}
\end{figure}

As shown in Refs. \cite{RSM,SRS}, the bound state contributions are obtained from the poles of Im$\Sigma(1,z)$ which cannot be neglected expanding the spectral function. A cluster decomposition of the self-energy has been proposed, see Fig. \ref{fig:clusterdecomp}.
The diagrams are calculated as 
\begin{equation}
\label{SigmaT}
 \Sigma (1, z_\nu) = \sum_{A>1} \sum_{\Omega_\lambda, 2...A} G^{(0)}_{A-1}(2,...,A,\Omega_\lambda - z_\nu) {\rm T}_A(1...A,1'...A',\Omega_\lambda)\,.
\end{equation}
The free $(A-1)$ quasiparticle propagator $ G^{(0)}_{A-1} $ and the  ${\rm T}_A$ matrix are given in App. \ref{app.0}.
The ${\rm T}_A$ matrices are related to the $A$-particle Green functions which read in bilinear expansion
\begin{equation}
G_A(1...A,1'\dots A',z_A)=\sum_{\nu P}\psi_{A \nu P}(1\dots A)
\frac{1}{ z_A-E^{\rm qu}_{A, \nu}(P)} \psi^*_{A \nu P}(1'\dots A')\,.
\label{bilinear}
\end{equation}
The $A$-particle wave function $\psi_{A \nu P}(1\dots A) $
and the corresponding eigenvalues $E^{\rm qu}_{A, \nu}(P)$ result from solving the in-medium
Schr\"odinger equation, see the following Sec. \ref{sec2.2}.  Besides the bound states, the summation over the internal quantum states $\nu$ includes also the scattering states.

The evaluation of the equation of state in the low-density limit is
straightforward, see App. \ref{app.0}.
Considering only the bound-state contributions, we obtain the result 
\begin{eqnarray}
  n_p(T,\mu_p,\mu_n)&=& \frac{1 }{ \Omega} \sum_{A,\nu,P}Z 
f_{A,Z}[E^{\rm qu}_{A,\nu}(P;T,\mu_p,\mu_n)]\,, \nonumber\\  n_n(T,\mu_p,\mu_n)&=& \frac{1}{ \Omega} \sum_{A,\nu,P}(A-Z) 
f_{A,Z}[E^{\rm qu}_{A,\nu}(P;T,\mu_p,\mu_n)]\,
\label{quasigas}
\end{eqnarray}
for the EoS describing a mixture of components (cluster quasiparticles) obeying Fermi or Bose
statistics. The NSE is obtained in the low-density limit if the in-medium energies  $E^{\rm qu}_{A,\nu}(P;T,\mu_p,\mu_n)$ can be replaced by the binding energies of the isolated nuclei $E^{(0)}_{A,\nu}(P)$.
Note that at low temperatures Bose-Einstein condensation
may occur.  

We discuss the cluster decomposition of the single-nucleon self-energy, Fig. \ref{fig:clusterdecomp}, in comparison with other approximations. Restricting the cluster decomposition only to the contribution of two-particle correlations, we obtain the so-called $T_2G$ approximation. In this approximation, the Beth-Uhlenbeck formula is obtained for the EoS, as shown in \cite{RSM,SRS}. The relation between this T matrix approach and the Brueckner G matrix approach was discussed in detail in Ref. \cite{ARSKK}. Extended work has been performed using sophisticated interaction potentials  to evaluate the quasiparticle energies in the DBHF approximation, for recent reviews see Refs. \cite{fuc06,fuchs,mar07}.

Replacing the $T_2$ matrix in Born approximation with the interaction potential $V$, we obtain the Hartree-Fock approximation 
\begin{equation}
\Delta^{\rm HF}(1) = \sum_2[V(12,12)-V(12,21)] f_{1,\tau_2}[E(2)+\Delta^{\rm HF}(2)]\,.
\end{equation}
In this approximation, all correlations in the medium are neglected. The self-energy does not depend on frequency, i.e. it is instantaneous in time, with vanishing imaginary part.

\subsection{Effective wave equation for the $A$-nucleon cluster}
\label{sec2.2}

We consider the propagation of an  $A$-nucleon cluster in warm, dense matter which is described by the $A$-particle Green function $G_A$. The solution will not only enter the cluster decomposition of the single-nucleon self-energy as considered in Sec. \ref{sec2.1} to calculate the contribution of bound states to the nuclear matter EoS in a systematic way. It determines also the cluster decomposition of other quantities such as the polarization function, the dynamical structure factor etc. We can proceed as above in the single-nucleon case and investigate higher order correlation functions. The $A$-nucleon spectral function $S_A$ can be introduced which is related to $G_A$. Cluster quasiparticle excitations are determined by $\delta$-like peaks in the spectral function $S_A$. 

For the $A$-particle Green function, the perturbation expansion can be represented by Feynman diagrams. New elements such as the $A$-particle self-energy can be introduced, see Ref. \cite{KKER,RSM,cmf}. In the low-density limit, the bound and scattering states of the $A$-particle cluster are obtained performing the partial summation of ladder diagrams.

The interaction of the $A$-nucleon cluster with the surrounding nucleons can be considered in the cluster-mean field approximation given in  App. \ref{app.1}. Since nuclear matter can form clusters, the interaction of the $A$-nucleon cluster under consideration is taken with an arbitrary cluster $B$ of the surrounding matter in first order of the nucleon-nucleon interaction, but full antisymmetrization of all nucleons in the cluster $A$ and $B$. As a result, not only the Hartree shift due to the interaction with the surrounding nucleons in free as well as in cluster states is obtained. Also the the Pauli blocking terms due to the occupation of the phase space by free nucleons as well as nucleons which are bound in clusters is taken into account. In higher orders, the Bose-enhancement or Pauli blocking of nucleonic clusters is obtained.

At present, the full self-consistent solution of the cluster-mean field approximation is out of reach. Only special cases can be solved such as $\alpha$ cluster nuclei \cite{THSR} where an  $\alpha$ cluster is considered in a surroundings consisting of a few $\alpha$ clusters, with full antisymmetrization on the nucleonic level.

We calculate the modification of the $A$-nucleon cluster due to the surroundings considering only self-energy and Pauli-blocking terms in the cluster-mean field approximation. Correlations in the medium are neglected. In momentum space, the distribution of free nucleons, forming a Fermi sphere, and the bound states which are characterized by a wave function, are different functions, but the total amount of the occupied phase space is determined by the total density of the protons or neutrons. Considering the uncorrelated medium, the phase space occupation is described by a Fermi distribution function normalized to the total density of nucleons, see App. \ref{app.1},
\begin{equation}
\label{f1}
\tilde f_1(1) = \frac{1}{\exp[E_1^{\rm
qu}(1)/T- \tilde \mu_\tau/T] +1} \approx \frac{n_\tau}{2} \left(\frac{2 \pi \hbar^2}{m T}\right)^{3/2} 
e^{-E_1^{\rm qu}(1)/T}
\end{equation}
in the low-density, non-degenerate limit ($\tilde \mu_\tau <0  $). The effective chemical potential  $\tilde \mu_\tau$ is
determined by the normalization condition $2 \Omega^{-1} \sum_p \tilde f_1(p) =
n_{\tau}$, where $\tau $ denotes isospin (neutron or proton), and has to be expressed in terms of these densities.

In ladder approximation, the $A$-particle Green function obeys a Bethe-Salpeter equation (BSE)
\begin{eqnarray}
&&G_A(1...A,1'\dots A',z_A)=G^{(0)}_A(1...A,z_A)
\delta_{11'}\dots\delta_{AA'} \nonumber \\ && + \sum_{1''\dots A''}G^{(0)}_A(1...A,z_A) 
V_A(1...A,1''\dots A'')G_A(1''...A'',1'\dots A',z_A)\,,
\label{BSE}
\end{eqnarray}
where $ V_A(1...A,1'\dots A')= \sum_{i<j}V(ij,i'j')\prod_{k \neq
  i,j} \delta_{kk'}$ is the interaction within the $A$-particle
cluster. $z_A$ denotes a fermionic or bosonic Matsubara frequency.
The free $A$-quasiparticle Green function results as
\begin{equation}
G^{(0)}_A(1...A,z_A)=\frac{[1-\tilde f_1(1)]\dots [1-\tilde f_1(A)]-\tilde f_1(1)\dots \tilde f_1(A) }{
z_A - E^{\rm qu}_1(1)-\dots- E_1^{\rm qu}(A)}\,.
\label{freeGF}
\end{equation}

The solution of the BSE is given by the bi-linear expansion (\ref{bilinear}). The $A$-particle wave function 
and the corresponding eigenvalues follow from solving the in-medium
Schr\"odinger equation  
\begin{eqnarray}
&&[E_1^{\rm qu}(1)+\dots + E_1^{\rm qu}(A) - E^{\rm qu}_{A \nu}(P)]\psi_{A \nu P}(1\dots A)
\nonumber \\ &&
+\sum_{1'\dots A'}\sum_{i<j}[1-\tilde f_1(i)- \tilde f_1(j)]V(ij,i'j')\prod_{k \neq 
  i,j} \delta_{kk'}\psi_{A \nu P}(1'\dots A')=0\,.
\label{waveA}
\end{eqnarray}
This equation contains the effects of the medium in the quasiparticle shift 
as well as in the Pauli blocking terms. 
Obviously the bound state wave functions and energy eigenvalues as
well as the scattering phase shifts become dependent on temperature
and density. Two effects have to be considered: the single-nucleon quasiparticle
energy shift and the Pauli blocking. 

Detailed investigations of Eq. (\ref{waveA}) have been performed for $A=2$, see Ref. \cite{SRS}, describing interesting physics.
From the solution of this in-medium two-particle Schr\"odinger
equation or the corresponding T matrix, the medium dependent scattering and possibly
bound states are obtained, as well as the formation of a quantum condensate, including the crossover from Bose-Einstein condensation to Cooper pairing. Due to the self-energy shifts and the Pauli
blocking, the energy of the deuteron, $E^{\rm qu}_d(P;T,\mu_p, \mu_n)$, as
well as the scattering phase shifts, $\delta_{\tau}(E,P;T,\mu_p,
\mu_n)$ ($\tau$ denoting the isospin singlet or triplet channel), will depend on the temperature and the chemical potentials. For a separable interaction $V(12,1'2')$ like the PEST4
potential \cite{PEST}, solutions of the in-medium two-particle Schr\"odinger
equation can be found, e.g., in Refs. \cite{SRS,ARSKK,SSAR}.
We will evaluate the medium shift of the binding energy in the following section using perturbation
theory.

\section{Expressions for the light nuclei quasiparticle shifts}
\label{sec.3}

\subsection{Solution of the few-nucleon effective wave equation}
\label{subsec.3.1}

The quasiparticle energies in warm and dense nuclear matter, $ E^{\rm qu}_{A \nu}(P;T,n_p,n_n)$, are well defined functions of temperature and total proton and neutron densities, given by a peak in the $A$-particle spectral function.  To evaluate these quantities for infinite matter from a first-principle quantum statistical approach, we have to perform some approximations.  Note that the quasiparticle shifts can also be introduced phenomenologically fitting empirical data. This is well-known for the single-nucleon quasiparticle energy which can be adapted to reproduce the structure of nuclei \cite{Typel}.

The solution of the few-body in-medium Schr\"odinger equation (\ref{waveA}) for separable interaction is simple in the case $A$=2. For $A$=3, 4, a Faddeev approach can be used, see \cite{Bey,Armen}. To obtain explicit expressions for the quasiparticle energy shifts, we will apply perturbation theory, which can be justified in the low-density region. Denoting the unperturbed wave function of the $A$-nucleon cluster with $\varphi_{A \nu P}(1\dots A)$, we have
\begin{equation}
\label{pertA}
E^{\rm qu}_{A \nu}(P) \approx \sum_{1...A,1'...A'} \varphi_{A \nu P}(1\dots A) H^{\rm eff}(1\dots A,1'\dots A') \varphi_{A \nu P}(1'\dots A')\,,
\end{equation}
where the form of $ H^{\rm eff}(1\dots A,1'\dots A') $ is given by Eq. (\ref{waveA}) after symmetrization. 

Before calculating the in-medium quasiparticle energy eigenvalues (\ref{pertA}), $A \le 4$, we first consider the isolated A-nucleon problem to determine the unperturbed wave function $\varphi_{A \nu P}(1\dots A)$. Extended analysis has been carried out by Wiringa et al. using Green's function Monte Carlo calculations \cite{Wir01}, where the AV18/UIX potential was used. Some properties are given in Tab. \ref{tab:table1}. There is excellent agreement between theory and experimental data \cite{atomdata}.

\begin{table}
\caption{\label{tab:table1} Light cluster properties at zero density}
\begin{ruledtabular}
\begin{tabular}{|l|l|l|c|l|l|}
  & binding & mass & spin & rms-radius & rms-radius
  \\
  & energy&  &  &  (charge)& (point)\\ 
&[MeV] &[MeV/$c^2$] & & [fm] &[fm]  \\
\hline
$n$ & 0 & 939.565  & 1/2&0.34  & 0  \\
$p$ ($^1$H) & 0 & 938.783  & 1/2 & 0.87  & 0 \\
$d$ ($^2$H) & -2.225 & 1876.12  & 1 & 2.14  & 1.96 \\
$t$ ($^3$H) & -8.482  & 2809.43  & 1/2 & 1.77  & 1.59   \\
$h$ ($^3$He) & -7.718  & 2809.41  & 1/2 & 1.97  & 1.76   \\
$\alpha$ ($^4$He) & -28.30  & 3728.40  & 0 & 1.68   & 1.45  \\
\end{tabular}
\end{ruledtabular}
\end{table}

In contrast to the second virial coefficient of the EoS given below in Sec. \ref{sec:BU}, which is determined by on-shell properties (binding energy and scattering phase shifts of the two-nucleon problem) determined from experiments, the nucleon interaction potential and the bound state wave function must be known to evaluate the quasiparticle energies . This potential is derived fitting empirical data which have to be reproduced. In previous work \cite{SRS}, the PARIS potential \cite{PEST} has been used. Alternatively, the AV18/UIX potential \cite{Wir01} can be taken. Here, we will use simple separable interaction potentials, which are fitted to the binding energy and the rms radius value of the respective cluster, in order to obtain  analytical expressions for the quasiparticle shifts as function of $P, T, n_p, n_n$. Details are given in App. \ref{app.2}.

\subsubsection{Gaussian wave function approach}

For the separable pair interaction in the $A$-nucleon cluster a Gaussian form is taken,
\begin{equation}
\label{Gausspot}
V_\nu(12,1'2') = \lambda_\nu e^{-\frac{(\vec p_2 -\vec p_1)^2}{4 \gamma_\nu^2}} e^{-\frac{(\vec p'_2 -\vec p'_1)^2}{4 \gamma_\nu^2}} \delta_{p_1+p_2,p'_1+p'_2} \delta_{\sigma_1,\sigma_1'}\delta_{\sigma_2,\sigma_2'}\delta_{\tau_1,\tau_1'}\delta_{\tau_2,\tau_2'}\,, 
\end{equation}
where $\nu = \{A,Z\}:=\{d,t,h,\alpha\}$ denotes the cluster under consideration.

To solve the $A$-particle Schr\"odinger equation, a variational approach is used. Two different classes of functions are considered. First, a Gaussian wave function is taken which allows for analytical expressions in evaluating the shifts. In the following Sec. \ref{subsub3.1.2} a better, but more complex Jastrow ansatz is considered which, however, allows in general only for numerical evaluation of the shifts.

For Gaussian wave functions, the center of mass motion can be easily separated. For vanishing center-of-mass motion, $P=0$, we have
\begin{eqnarray}
\label{varGauss}
\varphi^{\rm Gauss}_{\nu}(p_1...p_A)&=& \frac{1}{\rm norm_\nu}\,\, e^{-(p_1^2+...+p_A^2)/B_\nu^2}
\delta_{\vec p_1+...+\vec p_A,0} \,,
\end{eqnarray}
with the normalization $\sum_{p_1...p_A} |\varphi^{\rm Gauss}_{\nu}(p_1...p_A)|^2 = 1$.
The parameter $B_\nu$
 is fixed by the nucleonic point rms radius $\sqrt{ \langle r^2 \rangle_{\nu}}$, given in Tab. \ref{tab:table1},
\begin{equation}
\label{B}
B_\nu^2 = \frac{3 (A-1)}{A \langle r^2 \rangle_{\nu}},
\end{equation}
see App. \ref{app.2}. Values for $  B_\nu$ are presented in Tab. \ref{Tab.2a}.

Next we are interested in the parameter values $\tilde \lambda_\nu, \tilde \gamma_\nu$ of the potential (\ref{Gausspot}) which yield the binding energy $E^{(0)}_\nu$ of the nucleus as well as the nucleonic point rms radius, given in Tab. \ref{tab:table1}, using the variational ansatz (\ref{varGauss}). Calculating the kinetic energy as well as  the potential energy for a Gaussian wave function with range parameter $B_\nu$, we have
\begin{equation}
E^{(0)}_\nu =  \frac{3 (A-1)}{ 8}  \frac{\hbar^2 }{ m} B_\nu^2 + \frac{A (A-1)}{2} \lambda_\nu \frac{\gamma_\nu^6 B_\nu^3 }{
  \pi^{3/2} (B_\nu^2+2 \gamma_\nu^2)^3} \,.
\end{equation}
Varying $B_\nu$, the potential parameters which have the minimum energy $E^{(0)}_\nu$ at $B_\nu$, consistent with the rms radius value, are shown in Tab. \ref{Tab.2a}.

\begin{table}
\caption{ Light cluster wave function parameter at zero density from the Gaussian approach}
\begin{ruledtabular}
\begin{tabular}{|l|l|l|c|}
 $\nu$ & $\tilde \lambda_\nu$ & $\tilde \gamma_\nu$  & $B_\nu$ \\
&[MeV fm$^{3}$] & [fm$^{-1}$] & [fm$^{-1}$]   \\
\hline
$d$ ($^2$H) &-3677.2 & 0.753  & 0.625    \\
$t$ ($^3$H) & -1670.0  & 1.083  & 0.889     \\
$h$ ($^3$He) & -1957.5  & 0.960  & 0.804    \\
$\alpha$ ($^4$He) & -1449.6  & 1.152  & 1.034     \\
\end{tabular}
\end{ruledtabular}
\label{Tab.2a}
\end{table}

\subsubsection{Jastrow wave function approach}
\label{subsub3.1.2}

Of course, the variational solution of the separable Gauss potential (\ref{Gausspot}) by using Gaussians is not optimal,
which is clearly seen for the two-nucleon system where the exact solution is known. A Jastrow ansatz which reproduces this exact solution for $A=2$ is given by
\begin{equation}
\label{varphi}
\varphi^{\rm Jastrow}_\nu(\vec p_1\dots \vec p_A) = \frac{1}{N_\nu}\prod_{i<j}\frac{e^{-\frac{(\vec p_j-\vec p_i)^2}{4a_\nu^2}}}{\frac{(\vec p_j-\vec p_i)^2}{4 b_\nu^2 }+1}\,.
\end{equation}
The prefactor $N_\nu$ is determined by the normalization condition. The results for the parameter values $\lambda_\nu, \gamma_\nu$ of the interaction potential, which reproduce the binding energies and rms values of the cluster (see Tab. I), as well as the parameter values $a_\nu, b_\nu$ characterizing the wave function (\ref{varphi}) within the variational approach, are given in Tab. \ref{Tab.2}. For the calculations see App. \ref{app.2}.

\begin{table}
\caption{\label{tab:table2} Light cluster wave function parameter at zero density from the Jastrow approach}
\begin{ruledtabular}
\begin{tabular}{|l|l|l|c|l|}
 $\nu$ & $\lambda_\nu$ & $\gamma_\nu$ &$ a_\nu$ & $b_\nu$ 
  \\
&[MeV fm$^{3}$] & [fm$^{-1}$] & [fm$^{-1}$] & [fm$^{-1}$]   \\
\hline
$d$ ($^2$H) &-1287.4 & 1.474  &  1.474 &  0.2317  \\
$t$ ($^3$H) & -1283.8  & 1.259  & 1.742 & 0.592    \\
$h$ ($^3$He) & -1527.0  & 1.105  & 1.528 & 0.542   \\
$\alpha$ ($^4$He) & -1272.9  & 1.231  & 2.151 & 0.912    \\
\end{tabular}
\end{ruledtabular}
\label{Tab.2}
\end{table}

\subsection{Expressions for the quasiparticle shifts}
\label{subsec.3.2}

\subsubsection{Nucleon quasiparticles}
\label{subsub3.2.1}

The in-medium single-nucleon dispersion relation $E^{\rm qu}_{1}(P)$ can be expanded for small momenta $P$ as 
\begin{equation}
\label{quasinucleonshift}
E^{\rm qu}_{1}(P)=  \frac{\hbar^2}{2 m_1}P^2 + \Delta E^{\rm SE}_{1}(P)=\Delta E^{\rm SE}_{1}(0) + \frac{\hbar^2}{2 m_1^*}P^2 + {\mathcal O}(P^4)\,,
\end{equation}
where the quasiparticle energies are shifted by $\Delta E^{\rm SE}_{1}(0)$, and $m_1^*$ denotes the effective mass of  neutrons ($\tau_1=n$) or protons ($\tau_1=p$). Both quantities,  $\Delta E^{\rm SE}_{1}(0)$ and $m_1^*$, are functions  of $T,n_p,n_n$ characterizing the surrounding matter.

Different expressions are used to parametrize the nucleon quasiparticle shift at subsaturation density. The Skyrme parametrization \cite{Vau72} of $\Delta E^{\rm SE}_{\tau}(0)$, see App. \ref{app.2}, is used in a standard approach to the nuclear matter EoS by Lattimer and Swesty \cite{LS}. Different improvements have been performed to optimise the calculation for various nuclei. Alternatively, relativistic mean-field (RMF) approaches have been developed starting from a model Lagrangian which couples the nucleons to mesons.
The relativistic quasiparticle energy is given by
\begin{equation}
\label{RMFquasi}
E_\tau^{\rm qu}(P) = \sqrt{ \left[ m_\tau c^2-S(n_n,n_p,T) \right]^2+\hbar^2 c^2 P^2} 
+ V_\tau(n_n,n_p,T)  \,.
\end{equation}
Expressions for $S$ and $V_\tau$ can be given \cite{shen} determining the nucleon quasiparticle shift and the effective mass, which aim to obtain the nuclear matter EoS for supernovae collapses.  
Recent work on RMF parametrization \cite{Typel,fuc06} intend to reproduce properties of nuclei, but are also in agreement with microscopic DBHF calculations.

The EoS of asymmetric nuclear matter has been investigated in the low-density region below the nuclear saturation density \cite{mar07}, and expansions of the quasiparticle shift in powers of the density have been given.
The  Dirac mass and the Landau mass are considered \cite{fuchs}. A more detailed discussion of the nucleon quasiparticle approach which is very successful in describing nuclear matter near saturation density cannot be given here.

\subsubsection{Deuteron quasiparticles}
\label{subsub3.2.2}

In the low-density limit, the shift of the deuteron binding energy can be calculated from the
in-medium Schr\"odinger equation (\ref{waveA}) taking the medium modifications as correction.
Within perturbation theory, the shifts of the solution of this medium modified wave equation are given by the self-energy term and the Pauli blocking term,
\begin{equation}
\label{deldeut}
E^{\rm qu}_{d}(P)=E_d^{(0)}+ \frac{\hbar^2}{2 m_d}P^2+\Delta E_d^{\rm SE}(P)
 +\Delta E_d^{\rm Pauli} (P)\,,
\end{equation}
$m_d \approx 2m$ is the deuteron rest mass. After separation of the center-of mass motion, the free deuteron wave function $\varphi_d(\vec q_1)$  is taken as a function of the relative momentum $\vec q_1 = (\vec p_2 - \vec p_1)/2$ (neglecting the proton-neutron mass difference).  Perturbation theory gives for the self-energy term due to the single-nucleon self-energy shift $\Delta E_1^{\rm SE}(P) = E_1^{\rm qu}(P)-E(1)$ in the medium part of the effective Hamiltonian  (\ref{waveA})
\begin{equation}
\Delta E_d^{\rm SE}(P)= \frac{1}{N_d}\sum_{\vec q_1} \varphi^*_d(\vec q_1) \left[ \Delta E_n^{\rm SE}\left(\frac{\vec P}{2} +\vec q_1\right)+\Delta E_p^{\rm SE}\left(\frac{\vec P}{2} -\vec q_1\right) \right] \varphi_d(\vec q_1).
\end{equation}
$N_d=\sum_{\vec q_1} | \varphi_d(\vec q_1) |^2$ is the normalization.
For a separable interaction, the Pauli blocking term reads
\begin{equation}
\label{Paulid}
\Delta E_d^{\rm Pauli}(P) =  -\frac{1}{N_d}\sum_{\vec q_1,\vec q_1{}'} \varphi^*_d(\vec q_1) \left[ \tilde f_n\left(\frac{\vec P}{2} +\vec q_1\right)+\tilde f_p\left(\frac{\vec P}{2} -\vec q_1\right) \right] V(\vec q_1,\vec q_1{}')
\varphi_d(\vec q_1{}')\,,
\end{equation}
where the distribution function $\tilde f_1(\vec p)$ is given by Eq. (\ref{f1}).
After applying the unperturbed Schr\"odinger equation, the interaction in Eq. (\ref{Paulid}) can be eliminated. With the reduced mass $m/2$ we obtain
\begin{equation}
 \label{deltapauli}
\Delta E_d^{\rm Pauli}(P) = \frac{1}{N_d}\sum_{\vec q_1} |\varphi_d(\vec q_1)|^2 \left[ \tilde f_n\left(\frac{\vec P}{2} +\vec q_1\right)+\tilde f_p\left(\frac{\vec P}{2} -\vec q_1\right) \right] \left( \frac{\hbar^2}{m} q_1^2- E_d^{(0)} \right)\,.
\end{equation}

We give some more explicit expressions. Similar to the effective mass representation for the single-nucleon quasiparticle states, we can introduce the deuteron shift and effective mass according to 
\begin{equation}
\label{deldqu}
E^{\rm qu}_{d}(P)=E_d^{(0)}+\Delta E_d^{}(0)+ \frac{\hbar^2}{2 m^*_d}P^2+{\mathcal O}(P^4)\,.
\end{equation}
The self-energy contribution to the deuteron shift contains the contribution of the single-nucleon shift
\begin{equation}
\label{deldrigid}
\Delta E_d^{\rm rigid \,\, shift}(0)=\Delta E_n^{\rm SE}(0)+\Delta E_p^{\rm SE}(0)\,.
\end{equation}
The rigid shift of the nucleons is also present in the continuum states and will not change the binding energy of the bound states. It can be incorporated in the chemical potentials $\mu_n, \mu_p$, similar to the rest mass of the nucleons.

If we consider the effective mass of the nucleons $m_\tau^* \neq m_\tau$, we get the contribution 
\begin{equation}
\label{deldeffmass}
\Delta E_d^{\rm eff. mass}(P)= \frac{1}{N_d}\sum_{\vec q_1} |\varphi_d(\vec q_1)|^2 \frac{\hbar^2}{2 m} \left(\frac{m_p}{m_p^*}+\frac{m_n}{m_n^*}-2\right) \left(q_1^2 + \frac{P^2}{4} \right)\,.
\end{equation}
The first term contributes to the deuteron shift $\Delta E_d^{\rm SE}(0)$, the last term ($\propto P^2$) to the deuteron effective mass. With $\Delta m_d=m_d^*-m_d$ we find for the self-energy contribution
\begin{equation}
\frac{\Delta m_d^{\rm SE}}{m_d}=-\frac{1}{8}  \left(\frac{m_p}{m_p^*}+\frac{m_n}{m_n^*}-2\right)\,.
\end{equation}

For the further evaluation, we need the free deuteron wave function. For the solution (\ref{varphi}) of the Gaussian interaction (\ref{Gausspot}), used in the Jastrow ansatz with parameter values given in Tab. \ref{Tab.2}, we obtain for the Pauli blocking shift (\ref{deltapauli}) the expression $\Delta E_d^{\rm Pauli,Jastrow}(P) = (n_p+n_n) \delta E_d^{\rm Pauli,J}(P)$ with
\begin{eqnarray}
 \label{deltapauligauss}
&&\delta E_d^{\rm Pauli,J}(P) = \frac{1}{2} \left( \frac{2 \pi \hbar^2}{m T} \right)^{3/2} 
\int dq q^2 \frac{e^{-2 q^2/a_d^2}}{(q^2/b_d^2+1)^2} e^{-\frac{\hbar^2}{2 m T} \left(\frac{P^2}{4}+q^2\right)} \\ && \times \frac{mT}{\hbar^2Pq} \left(e^{\frac{\hbar^2Pq}{2mT}}- e^{-\frac{\hbar^2Pq}{2mT}} \right)\left( \frac{\hbar^2}{m} q_1^2- E_d^{(0)} \right)
\frac{\sqrt{2} a_d}{\sqrt{\pi} b_d^4 \left(-1+e^{2\frac{b_d^2}{a_d^2}} \frac{ \sqrt{\pi}}{2 \sqrt{2}\frac{b_d}{a_d} } (1+ 4\frac{b_d^2}{a_d^2}) {\rm erfc}[\sqrt{2}\frac{b_d}{a_d}]\right)},\nonumber
\end{eqnarray}
taking the nucleons in the medium as non-degenerate what, at finite temperatures, is justified in the low-density limit. (In the degenerate case, the Boltzmann distribution has to be replaced by the Fermi distribution.) 
In particular, we find the quasiparticle shift at $P=0$
\begin{eqnarray}
 \label{deltapauligauss0}
&&\delta E_d^{\rm Pauli,J}(0) = \frac{1}{2} \left( \frac{2 \pi \hbar^2}{m T} \right)^{3/2} 
\frac{\hbar^2 a_d^2}{2m}\nonumber\\ && \times \frac{\left[1+ \frac{\hbar^2 a_d^2}{4 m T}\right]^{-1/2}-e^{2\frac{b_d^2}{a_d^2}\left(1+ \frac{\hbar^2 a_d^2}{4 m T}\right)} \frac{ \sqrt{2 \pi} b_d}{a_d }\,\, {\rm erfc}\left[\sqrt{2}\frac{b_d}{a_d} \sqrt{1+ \frac{\hbar^2 a_d^2}{4 m T}}\right]}{-1+e^{2\frac{b_d^2}{a_d^2}} \frac{ \sqrt{\pi} a_d}{2 \sqrt{2} b_d } \left(1+ 4\frac{b_d^2}{a_d^2}\right) \,\,{\rm erfc}\left[\sqrt{2}\frac{b_d}{a_d}\right]}\,.
\end{eqnarray}
Values for the deuteron quasiparticle shift are given below in Tab. \ref{Tab.3}

To obtain the Pauli blocking contribution to the deuteron effective mass, we can expand the Pauli blocking shift (\ref{deltapauligauss}) for small values of $P$. We will not give the corresponding expressions here. The effective mass approximation for the deuterons in warm dense matter is limited to small values of $P$.
For arbitrary $P$, after averaging over the direction between $\vec q_1$ and $\vec P$, we get the approximation
\begin{equation}
\Delta E_d^{\rm Pauli,Jastrow}(P) \approx \Delta E_d^{\rm Pauli,Jastrow}(0)\,e^{-\frac{\hbar^2}{8mT}P^2}\,.
\end{equation}
Together with Eqs. (\ref{deltapauligauss0}), (\ref{deldeffmass}), (\ref{deldrigid}), we obtain the momentum-dependent quasiparticle energy (\ref{deldeut}) describing the deuteron in matter.

In contrast to the self-energy shift, the Pauli blocking shift depends strongly on temperature by the following reasons. The binding energy per nucleon characterizes the extension of the wave function in momentum space. The form of the Fermi distribution function is determined by the temperature $T$ which we consider here to be of the same order as the binding energy per nucleon. Further, the Pauli blocking term is strongly depending on the center of mass momentum $P$ because of the overlap of the deuteron wave function in momentum space with the Fermi sphere. Therefore, the bound states with high momentum $P$ are less modified by the Pauli blocking effects. The evaluation of the deuteron quasiparticle energy is further improved using a better wave function, an appropriate interaction potential and avoiding perturbation expansions. Comparison with other potentials such as the Yamaguchi Lorentzian formfactor has been performed which give only small changes (below 10 \%) to the deuteron quasiparticle shift.

\subsubsection{Tritium and Helium in matter}
\label{subsub3.2.3}

We now consider the clusters with $A$=3 $(t,h)$ and $A$=4  $(\alpha)$. Some of the relations given for the deuteron case in Sec. \ref{subsub3.2.2} can be generalized to higher values of $A$.
In the low-density limit, the shift of the cluster binding energy can be calculated from the
effective Schr\"odinger equation (\ref{waveA}) taking the medium modifications as correction,
\begin{equation}
E^{\rm qu}_{\nu}(P)=E_\nu^{(0)}+ \frac{\hbar^2}{2 m_\nu}P^2+\Delta E_\nu(P)\,
\end{equation}
with $\nu=\{A,Z\} = \{t,h,\alpha\}$, and $m_\nu \approx A m$ denoting the rest mass of the cluster.  Within perturbation theory, the shift $\Delta E_\nu(P)= \Delta E_\nu^{\rm SE}(P) +\Delta E_\nu^{\rm Pauli} (P)$ consists of the self-energy and Pauli-blocking term which can be calculated from the effective wave equation  (\ref{waveA}) with the unperturbed wave function $\varphi_\nu(\vec p_1,\dots,\vec p_A)$. After separation of the center-of mass motion, $\varphi_\nu(\vec q_1,\dots,\vec q_{A-1})$ is a function of the remaining Jacobian momenta $\vec q_i$, see App. \ref{app.2}. We find for the energy shifts due to the single-nucleon self-energy shift in the wave equation  (\ref{waveA})
\begin{equation}
\Delta E_\nu^{\rm SE}(P)= \frac{1}{N_\nu}\sum_{\vec q_i} |\varphi_\nu(\vec q_i)|^2 \left[ \Delta E_n^{\rm SE}(\vec p_1)+\dots+ \Delta E_n^{\rm SE}(\vec p_A) \right] .
\end{equation}
and for the Pauli blocking term
\begin{equation}
\label{delApauli}
\Delta E_\nu^{\rm Pauli}(P) =  -\frac{1}{N_\nu}\sum_{\vec q_i,\vec q_i{}'} \varphi^*_\nu(\vec q_i) 
\sum_{i<j} \left[\tilde f_1(p_i)+ \tilde f_1(p_j)\right] V(p_i p_j ,p_{i}' p_{j}')\prod_{k \neq 
  i,j} \delta_{p_k, p_{k}'}
\varphi_\nu({\vec q_i}{}')\,,
\end{equation}
where $N_\nu=\sum_{\vec q_i} | \varphi_\nu(\vec q_i) |^2$ is the normalization. The single-nucleon momenta $\vec p_i$ have to be expressed by Jacobi momenta $\vec q_i$ including the center of mass momentum $\vec P$.
The Pauli blocking term results as
where also the momenta $\vec p_i$ have to be expressed in terms of the Jacobi momenta $\vec q_i$.

As for the single-nucleon and deuteron states, we can introduce the quasiparticle shift and effective mass according to 
\begin{equation}
\label{delAqu}
E^{\rm qu}_{\nu}(P)=E_\nu^{(0)}+\Delta E_\nu^{}(0)+ \frac{\hbar^2}{2 m^*_\nu}P^2+{\mathcal O}(P^4)\,.
\end{equation}
The self-energy contribution to the quasiparticle shift contains the contribution of the single-nucleon shift
\begin{equation}
\label{delArigid}
\Delta E_\nu^{\rm rigid \,\, shift}(0)= (A-Z) \Delta E_n^{\rm SE}(0)+ Z \Delta E_p^{\rm SE}(0)\,.
\end{equation}
 Evaluating the distribution functions $f_{A,Z}[E^{\rm qu}_{A,\nu}(P)]$, Eq. (\ref{quasigas}), this contribution to the quasiparticle shift can be included renormalizing the chemical potentials $\mu_n, \mu_p$.
The further discussion of the self-energy contribution to the quasiparticle shift and effective mass can be performed in analogy to the deuteron case and will not be repeated here.

Whereas in the deuteron case it was possible to eliminate the interaction potential calculating the Pauli blocking shift, Eq. (\ref{deltapauli}), this is not longer possible in the case $A=3,4$, and we have to evaluate expression (\ref{delApauli}).
Analytical expressions for the Pauli blocking shift will be given in the following Sec. \ref{quasiparticle shifts}, based on the more simple Gaussian ansatz for the unperturbed wave function of the cluster.
For the Jastrow ansatz (\ref{varphi}), results for $ \delta E_\nu^{\rm Pauli,J}(0) $ are given below in Tab. \ref{Tab.3}.

As already discussed for the deuteron case, after angular averaging the Pauli blocking shift can be approximated as
\begin{equation}
\label{delpauli0P}
\Delta E_\nu^{\rm Pauli}(P) \approx \Delta E_\nu^{\rm Pauli}(0) \, e^{-\frac{\hbar^2 P^2}{2 A^2 m T}}\,.
\end{equation}
It leads to the dissolution of the bound states below saturation density, starting at the Mott baryon density $n_B^{\nu,\rm Mott}(T,\alpha)$ depending on temperature $T$ and asymmetry parameter $\alpha$, where the $A$-nucleon bound state with $P=0$ merges in the continuum of scattering states \cite{RSM}.

\section{Results for the cluster quasiparticle shifts at low densities}
\label{quasiparticle shifts}

We collect some results for the quasiparticle energy 
\begin{equation}
\label{finalshift}
E^{\rm qu}_{\nu}(P;T,n_n,n_p)
= E_{\nu}^{(0)}+\frac{\hbar^2P^2}{2 A m}+\Delta E_{\nu}^{\rm SE}(P)+\Delta E_{\nu}^{\rm Pauli}(P)
+\Delta E_{\nu}^{\rm Coul}(P)
\end{equation}
for the light elements $\nu = \{d, t, h, \alpha\}= \{ ^2{\rm H},^3{\rm H},^3{\rm He},^4{\rm He}  \}$  in warm, dense nuclear matter. In addition to the single-particle self-energy shift $\Delta E_{\nu}^{\rm SE}(P)$ and the Pauli blocking term 
$\Delta E_{\nu}^{\rm Pauli}(P)$,  the quasiparticle energy shift contains also the Coulomb shift 
$ \Delta E_{\nu}^{\rm Coul}(P) $ which will not be elaborated here. The remaining two contributions are calculated within perturbation theory. 

The quasiparticle self-energy shift $\Delta E_{\nu}^{\rm SE}(P)$ is caused by the self-energy shift $\Delta E_1^{\rm SE}(P)$ of the single-nucleon energies. For the $A$-nucleon system, it contributes to the bound state energies (nuclei) as well as to the energy of scattering states, in particular the edge of the continuum of scattering states, see Ref. \cite{KKER}. In addition to the rigid shift, the quasiparticle self-energy shift at zero momentum $\Delta E_{\nu}^{\rm SE}(0)=\Delta E_{\nu}^{\rm rigid\, shift}+ 3(A-1)\hbar^2 B^2_{\nu}(m-m^*)/(8m^2)$ contains  the contribution of the effective nucleon masses, calculated for Gaussian wave functions where $B_\nu$ is given by Eq. (\ref{B}). Similarly, the self-energy contribution to the cluster effective mass can be calculated.

The Pauli blocking contribution $\Delta E_{\nu}^{\rm Pauli}(P)$ may be taken in the approximation (\ref{delpauli0P}) so that we discuss only the cluster quasiparticle shift $\Delta E_{\nu}^{\rm Pauli}(0)$ here. 
The Hamiltonian describing the unperturbed cluster contains parameters for the interaction potential which are determined such that the correct binding energy and point rms radius of the nuclei, given in Tab. \ref{tab:table1}, are reproduced. Two different approximations are considered.

(i) The variational function to solve the isolated cluster Schr\"odinger equation is taken as a Gaussian wave function with the parameter  $B_\nu$. A Gaussian potential (\ref{Gausspot}) is constructed which reproduces the correct values for the binding energies and the cluster point rms radii. The corresponding values for $\tilde \lambda_\nu, \tilde \gamma_\nu$ are given in Tab. \ref{Tab.2a}.
Now, having the interaction and the wave function to our disposal, we can calculate the shift of 
the binding energies in the low-density region using perturbation theory. 
Expanding with respect to the densities $n_n, n_p$, we have
\begin{equation}
 \Delta E_\nu^{\rm Pauli}(0)= [(1+\alpha_\nu) n_n +(1-\alpha_\nu) n_p] \delta  E_\nu^{\rm Pauli}(T) + {\mathcal O}(n_B^2)\,
\end{equation}
with $\alpha_d = 0, \alpha_t = 1/3,\alpha_h = -1/3,\alpha_\alpha = 0$.
In the non-degenerate case  
($\mu_\tau < 0$), expressions for $ \delta  E_\nu^{\rm Pauli}(T)  $ are obtained in analytic form as
\begin{eqnarray}
 \delta E_\nu^{\rm Pauli,G}(T) & = &- \frac{A (A-1)}{4 \pi^{3/2}} \left(\frac{2 \pi \hbar^2}{m T}\right)^{3/2}
\frac{\tilde \lambda_\nu B_\nu^3 \tilde \gamma_\nu^6}{(B_\nu^2+2 \tilde \gamma_\nu^2)^{3/2}  (B_\nu^2+2 \tilde \gamma_\nu^2 +
\frac{\hbar^2 B_\nu^2}{c_\nu m T} (B_\nu^2+d_\nu \tilde  \gamma_\nu^2))^{3/2}}\,,\nonumber \\&& {}
\label{shift3,4}
\end{eqnarray}
where $c_d=2,\, c_t=c_h=24, \,c_\alpha=16;\,\,\, d_d=0,\, d_t=d_h=14,\, d_\alpha=10$.

(ii) The same procedure as in (i), only the Jastrow function (\ref{varphi}) is taken as variational ansatz for the wave function. The Gaussian potential  (\ref{Gausspot}) with parameter values $ \lambda_\nu, \gamma_\nu$, see Tab. \ref{Tab.2}, reproduces the correct values for the binding energies and the cluster point rms radii. This variational ansatz gives the exact solution (\ref{deltapauligauss0}) of the two-nucleon problem, so that it is expected to yield better results also for the higher clusters. However, no analytic expressions are obtained for $A>2$.

 In Tab. \ref{Tab.3} the comparison between the Gaussian ($ \delta E_\nu^{\rm Pauli,G}$) and the Jastrow ansatz  ($ \delta E_\nu^{\rm Pauli,J}$) is shown for different temperatures. The differences are small (below 5 \%) for $A=3,4$ so that the analytical expression (\ref{shift3,4}) can be used. The differences are larger for $A=2$ because the deuteron wave function is not well approximated by a Gaussian. However, for $A=2$ we can take the Jastrow result which is given analytically in Eq. (\ref{deltapauligauss0}).

\begin{table}
\caption{Temperature dependence of the first order Pauli blocking shift for $A$ = 2, 3, 4. Comparison between the Gauss (G) and the Jastrow (J) approach. $T$ in [MeV], $\delta E_\nu^{\rm Pauli}$ in [MeV fm$^3$].}
\begin{ruledtabular}
\begin{tabular}{|r|r|r|r|r|r|r|r|r|}
 $T$ & $\delta E_d^{\rm Pauli,G}$ & $\delta E_d^{\rm Pauli,J}$ & $\delta E_t^{\rm Pauli,G}$ & $\delta E_t^{\rm Pauli,J}$& $\delta E_h^{\rm Pauli,G}$ & $\delta E_h^{\rm Pauli,J}$& $\delta E_\alpha^{\rm Pauli,G}$ & $\delta E_\alpha^{\rm Pauli,J}$ \\ 
\hline
20   &   157.9 &   172.9  &   482.7 &   470.77 &   438.9 &   432.52  &   967.7 &   950.16 \\
15   &   228.1 &   235.4  &   651.7 &   628.95 &   604.0 &   588.70  & 1263.8 & 1237.72 \\
10   &   371.4 &   352.5  &   967.7 &   905.37 &   906.7 &   871.81  & 1749.1 & 1711.83 \\
9     &   418.8 &   389.2  & 1037.9 &   986.54 &   998.4 &   957.23  & 1884.2 & 1844.90 \\
8     &   477.1 &   433.3  & 1140.0 & 1081.19 & 1106.6 & 1058.08  & 2037.5 & 1996.84 \\
7     &   550.3 &   487.7  & 1260.0 & 1192.82 & 1236.0 & 1178.72  & 2212.9 & 2171.86 \\
6     &   644.4 &   556.2  & 1402.7 & 1326.29 & 1392.6 & 1325.31  & 2415.0 & 2375.56 \\
5     &   768.8 &   645.2  & 1574.4 & 1488.50 & 1585.4 & 1506.83  & 2649.8 & 2615.75 \\
4     &   939.0 &   766.2  & 1784.3 & 1689.66 & 1827.2 & 1736.97  & 2925.4 & 2903.94 \\
3     & 1182.8 &   941.0  & 2045.4 & 1945.59 & 2137.3 & 2037.73  & 3252.1 & 3259.11 \\
2     & 1553.9 & 1212.0  & 2377.0 & 2282.48 & 2546.2 & 2447.21  & 3644.7 & 3720.67 \\
1     & 2169.8 & 1756.9  & 2808.9 & 2747.92 & 3104.5 & 3038.96  & 4123.0 & 4434.30 \\
\end{tabular}
\end{ruledtabular}
\label{Tab.3}
\end{table}

The wave functions of the isolated clusters determine the quasiparticle shifts within perturbation theory, in particular the Pauli blocking shift. The correct reproduction of the rms radii of the nuclei is important in order to estimate the region in phase space, which is needed to form the bound state. This region, however, may already be occupied by nucleons of the medium, leading to the Pauli blocking contribution in the quasiparticle shift. Improvements in calculating the wave functions of the isolated nuclei are possible using more sophisticated potentials such as BONN, PARIS \cite{PEST} or AV18/UIX \cite{Wir95} and applying advanced methods for the solution of the few-body problem such as the Faddeev-Yakubovski approach \cite{Bey,Armen} or the Green's function Monte Carlo method \cite{Wir01}. For tritium, a comparison with calculations of the wave function by Wiringa has been performed, and reasonable agreement with the Jastrow ansatz used here has been found. Further improvements of our results for the quasiparticle shifts using more advanced approaches to the few-nucleon problem may be the subject of future considerations.

\section{Application: Generalized Beth-Uhlenbeck equation and cluster virial expansion}
\label{sec:BU}

The nuclear matter equation of state (EoS) is obtained from Eq. (\ref{eosspec}) after specifying the self-energy $\Sigma(1,z)$. Taking into account only two-particle correlations in ladder approximation (\ref{SigmaT}), $A=2$, we obtain the generalized Beth-Uhlenbeck formula \cite{RSM,SRS}
\begin{equation}
\label{n12}
n_B(T,\mu_p,\mu_n)=n_1(T,\mu_p,\mu_n)+n_2(T,\mu_p,\mu_n)\,.
\end{equation}
The single-quasiparticle contribution is $n_1=n^{\rm free}_n+n^{\rm free}_p$,
where $n^{\rm free}_\tau(T, \mu_n, \mu_p) = 2 (2 \pi)^{-3} \int d^3P\,
f_\tau(E^{\rm qu}_\tau(P))$ describes the free quasi-particle contributions of neutrons and
protons, see also Eq. (\ref{nqu}). The
two-particle contributions $n_2 = n_2^{\rm bound} + n_2^{\rm scat}$
contains the contribution of deuteron-like quasiparticles (spin factor 3)
\begin{equation}
\label{boundBU}
n_2^{\rm bound}(T,\mu_p,\mu_n)=3 \int_{P>P_d^{\rm
Mott}}\frac{d^3P}{ (2 \pi)^3}\,\, f_d(E^{\rm qu}_d(P;T,\mu_p, \mu_n))\,,
\end{equation}
with $f_d(E)=[e^{(E -\mu_p -
\mu_n)/T}-1]^{-1}$, and scattering states of the isospin singlet and
triplet channel $\tau_2$ (degeneration factor $\gamma_{\tau_2}$)
\begin{equation}
\label{scattBU}
n_2^{\rm scat}(T,\mu_p,\mu_n)= \sum_{\tau_2}\gamma_{\tau_2} \int \frac{d^3P
}{ (2 \pi)^3} \int_0^\infty \frac{dE
}{ 2 \pi}\,\,f_{\tau_2}(\Delta E_d^{\rm SE}(P)+E)
\sin^2 \delta^{\rm qu}_{\tau_2}(E,P) \frac{d }{ dE} \delta^{\rm qu}_{\tau_2}(E,P)\,\,.
\end{equation}
$\Delta E_d^{\rm SE}(P)=\Delta E_n^{\rm SE}(P/2)+\Delta E_p^{\rm SE}(P/2)$ is the shift of the continuum edge
(self-energies at momentum $P/2$), and $ \delta^{\rm qu}_{\tau_2}(E,P)$ denotes the in-medium two-nucleon scattering phase shift in the channel $\tau_2$ with relative energy $E$ and center of mass momentum $P$. The quasideuteron binding energy $E^{\rm qu}_d(P;T,\mu_p, \mu_n)$ depends on temperature and nucleon densities of the medium, expressions are given in Sec. \ref{subsub3.2.2}. The shift of the quasiparticle binding energy increases with nuclear matter density so that the bound state may merge with the continuum of scattering states (Mott effect). $P_d^{\rm Mott}(T,\mu_p, \mu_n) $ denotes the momentum $P$ where,  at given temperature and nucleon density, the binding
energy of the deuteron bound state vanishes. Above the Mott density defined by $P_d^{\rm Mott}=0 $, the integral over $P$ in the bound state contribution $n_2^{\rm bound}$ (\ref{boundBU}) is restricted to the region where bound states can exist.

The generalized Beth-Uhlenbeck formula (\ref{n12}) can be considered as a virial
expansion of the EoS and was first applied to nuclear matter in Ref. \cite{RSM}. Results using the PEST4 interaction potential are shown in Ref. \cite{SRS}. Due to the inclusion of medium effects such as Pauli blocking and self-energy, a smooth transition has been obtained from the low density limit, describing nuclear statistical equilibrium between nucleons and deuterons, to high densities, where the nucleon quasiparticle picture can be used. Neglecting the medium effects, the ordinary Beth-Uhlenbeck formula is recovered, where the single nucleon contribution contains the energy $E(1)=\hbar^2 P^2_1/2m$ instead of the quasiparticle energy of free nucleons. In the correlated density $n_2$, the free deuteron binding energy and scattering phase shifts enter the ordinary Beth-Uhlenbeck formula. Here, the term sin$^2 \delta(E)$ preventing double counting if quasiparticle energies are taken in $n_1$, is absent, see \cite{SRS}. Then, the evaluation of the second virial coefficient can be traced back to on-shell properties such as binding energy and scattering phase shifts. This has been performed recently by Horowitz and Schwenk \cite{HS} using directly the empirical data, instead an interaction potential which is constructed to reproduce these data. However, the ordinary Beth-Uhlenbeck formula is restricted to only low densities.

As also shown in Ref. \cite{RSM}, the contribution of higher clusters to the EoS can be included  after a cluster decomposition of the self-energy $\Sigma(1,z)$, Eq. (\ref{SigmaT}). Neglecting the contribution of the scattering states, a generalized form of the ordinary NSE is obtained, as given by Eqs. (\ref{quasigas}).
The quasiparticle shift of the nuclei can be calculated in cluster-mean field approximation, see Eq. (\ref{waveA}). Results for light clusters, $A \le 4$, have been given in the previous Sec. \ref{quasiparticle shifts}. Due to the medium dependence of the quasiparticle energies, the contribution of $A>1$ to the EoS (\ref{quasigas}) fades with increasing density so that a smooth transition from low densities to saturation density is obtained. 

The light nuclei have also been considered by Horowitz and Schwenk \cite{HS} where a cluster-virial expansion was discussed. The second virial coefficient of the virial expansion is generalized taking into account also cluster-cluster scattering phase shifts. To obtain rigorous results in a given order of density avoiding double counting, and to include medium effects, one has to pass over to a systematic quantum statistical approach. However, the cluster-virial expansion may be of value when special clusters contribute dominantly to the density.

In principle, the approach given here can be extended to arbitrary clusters $A>4$. Estimates of the quasiparticle shift due to the interaction of the heavier cluster with the surrounding nuclear matter have been given in Ref. \cite{RSM2}. For large proton number $Z$, Coulomb effects become important. Calculating the composition of warm dense nuclear matter, the mass fraction of nucleons bound in heavy clusters is increasing with  decreasing temperature and increasing density. This limits the region in the phase space where the EoS is determined only by clusters with $A \le 4$.

The perturbative treatment of the cluster quasiparticle energy shift is justified in the first order of density. Higher orders can be  determined when comparing with the full solution of the medium modified Schr\"odinger equation for the $A$-nucleon cluster (\ref{waveA}). A term which is of second order with respect to the densities has been considered in Ref. \cite{R1}. Another effect is the formation of quantum condensates due to the Bose distribution function occurring in the bound state contribution (\ref{boundBU}), but also in the scattering contribution (\ref{scattBU}) to the density. Degeneracy effects are included in the quantum statistical approach used in this work. However, the evaluation of the EoS including quantum condensates such as pairing or quartetting needs further consideration.

We shortly comment the question of thermal stability in connection with the EoS (\ref{quasigas}). It is well-known that symmetric matter at zero temperature becomes instable against phase separation below saturation density. The conditions of thermodynamic stability are directly related to the behavior of the chemical potential as a function of the densities and will not be detailed here. Consequently, the EoS (\ref{quasigas}) describes nuclear matter in thermal equilibrium only outside the instability region. Parameter values belonging to metastable states or unstable states may occur  in nonequilibrium situations or in inhomogeneous systems when a local density approach is considered. They have also to be considered when the Maxwell construction is performed to determine the region of instability in the phase diagram.

The liquid-gas like phase transition  in nuclear matter has been considered in the EoS of Lattimer and Swesty \cite{LS} and in the EoS of Shen {\em et al}. \cite{shen}. Including Coulomb interaction, an optimum spherical density profile in a Wigner-Seitz cell approach was determined which can be interpreted as representative for heavy nuclei, but also for droplet formation. The $\alpha$ particle as representative of light clusters has been included. To mimic the density effects for the $\alpha$ particle, both approaches used the concept of the excluded volume which cannot be rigorously derived in a quantum statistical approach. In contrast, Pauli blocking considered here defines a microscopic process for the medium effects of light clusters.

\section{Conclusions}

In the low density limit,  the nuclear statistical equilibrium (NSE) with binding energies of the isolated nuclei is obtained from the quantum statistical approach to nuclear matter. We consider only light clusters, $A \le 4$, but heavier cluster may become more important as temperature goes down and density increases. Thus, focusing on only light elements in the EoS restricts the temperature and density parameter to the region where the mass fraction of heavy elements is small, but our approach may also be extended to heavier nuclei, see Ref. \cite{RSM2}.

Deviation from the NSE  are due to medium effects, which become relevant once the baryon density exceeds 10$^{-4}$ fm$^{-3}$. $A$-nucleon correlation effects are described by the $A$-nucleon spectral function which defines the $A$-cluster quasiparticles. The dependence of the cluster quasiparticle energy on temperature and nucleon densities is approximated by analytical expressions, Eq. (\ref{deltapauligauss0}) for $A=2$ and Eq. (\ref{shift3,4}) for $A=3,4$, combined with the momentum dependence according to Eq. (\ref{delpauli0P}). Compared with more accurate numerical calculations, deviations are of the order of 5 \%. Analytical expressions for the cluster quasiparticle shifts are convenient for calculating the thermodynamic properties of nuclear matter in a large parameter range. The evaluation of the cluster-quasiparticle shifts is further improved considering more sophisticated potentials and wave functions as obtained, e.g., in Green's function Monte Carlo approaches. The values given in the present work are approximate estimations, similar  as the Skyrme or RMF approaches for the single-nucleon quasiparticle case.

With the shift of the quasiparticle energies, properties such as the EoS (\ref{quasigas}) can be determined in the subsaturation region. It is possible to interpolate between the low-density limit where the NSE is valid, and  the saturation density where the single nucleon quasiparticle picture can be applied. 

To go beyond the quasiparticle picture, the full $A$-nucleon spectral function $S_A$ should be explored. Instead of $\delta$-like quasiparticle structures,  $S_A$ accounts for weakly bound states as well as scattering phase shifts including  resonances consistently. The full solution of the cluster-mean field approximation would be an important step in this direction.  The construction of a nuclear matter EoS remains a challenging topic not only in the high-density region, see \cite{Klahn:2006ir}, but also in the low-density region where the concept of nuclear quasiparticles given here may be a valuable ingredient.

\begin{acknowledgments}

The author thanks R. B. Wiringa (Argonne) for discussions on few-nucleon properties and for providing data of the bound state wave function, the Compstar collaboration, in particular D. Blaschke, for intensive discussions on the EoS, and the Yukawa Institute (Kyoto, YIPQS International Molecule Workshop), particularly H. Horiuchi, for hospitality when completing this work.

\end{acknowledgments}

\appendix

\section{T-matrix approach to the self-energy}
\label{app.0}

A detailed derivation of the expressions for the self-energy and the EoS can be found in the literature, see, e.g., Refs. \cite{FW,KKER,RSM,SRS}. We give here only some relations, using the short notations $E_1 = E_1^{\rm qu}(1)$, $E_{A \nu P} = E_{A,\nu}^{\rm qu}(P)$.

In Eq. (\ref{SigmaT}), the free $(A-1)$ quasiparticle propagator is
\begin{equation}
\label{G0quasi}
G^{(0)}_{A-1}(2,...,A, z) = \frac{1}{z-E_2-...-E_A} \frac{f_{1,Z_2}(2)...f_{1,Z_A}(A)}{f_{A-1,Z_{A-1}}(E_2+...+E_A)}\,.
\end{equation}
The ${\rm T}_A$ matrices are related to the $A$-particle Green functions
\begin{eqnarray}
\label{TAmatrix}
&&{\rm T}_A(1\dots A, 1' \dots A', z) =  V_A(1\dots A, 1' \dots
A')\nonumber\\
&&+V_A(1\dots A, 1'' \dots A'')
G_A(1''\dots A'', 1''' \dots A''', z) V_A(1'''\dots A''', 1' \dots A')
\end{eqnarray}
with the potential $ V_A(1\dots A, 1' \dots A') = \sum_{i<j}
V(ij,i'j') \prod_{k \neq i,j} \delta_{k,k'}$, and subtraction of double
counting diagrams when inserting the T matrices into the
self-energy. The $A$-particle propagator obeys a BSE (\ref{BSE}) and is solved by the bilinear expansion (\ref{bilinear}).

The evaluation of the equation of state in the low-density limit is
straightforward. With
\begin{eqnarray}
&&{\rm T}_A(1\dots A, 1' \dots A', z)=  \\ && \sum_{\nu,P} \frac{(z-E_1-...-E_A) \psi_{A\nu P}(1\dots
  A)\psi^*_{A\nu P}(1'\dots A') (E_{A\nu P}-E_{1'}-...-E_{A'}) }{ z- E_{A\nu P}} \nonumber
\end{eqnarray}
we can perform the $\Omega_\lambda$ summation in Eq. (\ref{SigmaT}). We obtain the result
\begin{eqnarray}
&&\sum_{\Omega_\lambda} \frac{1}{\Omega_\lambda - z_\nu-E_2-...-E_A} \frac{(\Omega_\lambda-E_1-...-E_A) 
 (E_{A\nu P}-E_{1}-...-E_{A})}{\Omega_\lambda-E_{A\nu P}} =\\
&&f_{A-1}(E_2+...+E_A) 
\frac{z_\nu-E_1}{z_\nu+E_2+...+E_A- E_{A\nu P}}-f_A(E_{A\nu P})
\frac{E_1+...+E_A- E_{A\nu P}}{z_\nu+E_2+...+E_A- E_{A\nu P}}\,. \nonumber
\end{eqnarray}
Taking Im\,$\Sigma(1,z)$ and integrating the $\delta$ function arising from the pole in the
denominator, the leading term in density is $f_1(E_{A\nu P}-E_2-...-E_A) 
f_{A-1}(E_2+...+E_A) = f_A(E_{A\nu P})$.
Neglecting the contribution of the scattering states, we obtain the generalized form (\ref{quasigas}) of the NSE.

\section{The cluster-mean field approximation}
\label{app.1}

The cluster-mean field approximation  \cite{cmf} is inspired by the chemical picture
where bound states are considered as new species, to be treated on the same
level as free particles. We consider the propagation of an
$A$-particle cluster ($\{A,\nu,P\}$) in a correlated medium. The corresponding 
$A$-particle cluster self-energy is treated to first order in 
the interaction with the single particles as well as with 
the $B$-particle cluster states ($\{B,\bar \nu,\bar P\}$) in the medium. The $B$-clusters in the surrounding medium are distributed according  to Eq. (\ref{faz}).
Full anti-symmetrization between both clusters $A$ and $B$ has to be performed, in analogy to the Fock term in the single-nucleon case.

For the $A$-particle
problem, the effective wave equation reads
\begin{eqnarray}
&&[E(1)+ \dots E(A) - E_{A \nu P}] \psi_{A \nu P}(1 \dots A)
 + \sum_{1'\dots A'}
\sum_{i<j}^A V_{ij}^A(1\dots A, 1'\dots A')  \psi_{A \nu P}(1' \dots A')
\nonumber\\ && + \sum_{1'\dots A'}
V_{\rm matter}^{A,{\rm mf}}(1\dots A, 1'\dots A')  \psi_{A \nu P}(1'
\dots A') = 0 \,,
\end{eqnarray}
with $V_{12}^A(1\dots A, 1'\dots A') = V(12,1'2') \delta_{33'} \dots
\delta_{AA'}$.  The effective potential $V_{\rm matter}^{A,{\rm mf}}
(1\dots A, 1'\dots A')$ describes the influence of the nuclear
medium on the cluster bound states and has the form
\begin{equation}
V_{\rm matter}^{A,{\rm mf}}(1\dots A, 1'\dots A') = \sum_i \Delta
(i) \delta_{11'} \dots \delta_{AA'} + {\sum_{i,j}}' \Delta
V_{ij}^A(1\dots A, 1'\dots A') \,.
\end{equation}
The self-energy like contribution
\begin{eqnarray}
&&\Delta(1) = \sum_2 V(12,12)_{\rm ex} \tilde f(2)
\nonumber\\ &&
 -  \sum^\infty_{B=2}
\sum_{\bar \nu \bar P} \sum_{2 \dots B} \sum_{1' \dots B'} f_{B,\bar Z}(E_{B  \bar \nu  \bar P})
\sum_{i<j}^B V_{ij}^B(1\dots B, 1'\dots B') \psi_{B  \bar \nu  \bar
  P}(1 \dots B) 
 \psi^*_{B  \bar \nu  \bar P}(1' \dots B')\, \nonumber
\end{eqnarray}
contains the Hartree-Fock quasiparticle shift. The interaction-like contribution
\begin{eqnarray}
&&\Delta V^A_{12}(1\dots A, 1'\dots A') = - \delta_{33'} \dots \delta_{AA'} \Biggl\{\frac{1}{2} \left[ \tilde
  f(1) + \tilde f(1')\right] 
V(12,1'2') +\\
&&
+\sum_{B=2}^\infty \sum_{\bar \nu \bar P} \sum_{\bar 2 \dots \bar
  B} \sum_{\bar 
  2'\dots \bar B'} f_{B,\bar Z}(E_{B \bar \nu \bar P})   
\sum_j^B V_{1j}^B (1 \bar2' \dots \bar B', 1' \bar2 \dots
\bar B) 
\psi^*_{B \bar \nu \bar P} (2 \bar2 \dots \bar B) \psi_{B \bar \nu \bar P} (2' \bar2'
\dots \bar B') \Biggr\}\, \nonumber
\end{eqnarray}
accounts for the Pauli blocking of the potential. The quantity
\begin{equation}
\tilde f(1) = f_1(1) + \sum^\infty_{B=2} \sum_{\bar \nu \bar P} \sum_{2 \dots B}
f_{B,\bar Z}(E_{B  \bar \nu  \bar P}) |\psi_{B \bar \nu \bar P}(1 \dots B)|^2\,
\end{equation}
describes the effective occupation in momentum space. The bound states contribute according to their wave function and  probability distribution. 
Note that within the mean-field approximation, the effective potential 
$V_{\rm matter}^{A,{\rm mf}}$ remains energy independent,
i.e.\ instantaneous.  Besides the Hartree-Fock term and the Pauli blocking term, determined by the effective
occupation  $\tilde f(1)$, the additional terms in $\Delta V^A_{12}$
and $\Delta(1)$ account for antisymmetrization.

\section{Wave functions and shifts}
\label{app.2}

\subsection{Quasinucleon}

The nucleon quasiparticle energy shift (\ref{quasinucleonshift}) contains the nucleon quasiparticle shift $ \Delta E_\tau^{\rm SE}$ and the effective nucleon mass $m_\tau^*$.
We expand with respect to the baryon density,
\begin{eqnarray}
\Delta E_\tau(P;T,n_B,\alpha) &=& \delta E_\tau(P;T,\alpha) n_B + { \mathcal{O}}(n_B^2),\nonumber \\
\frac{m^*_\tau}{m_\tau}(T,n_B,\alpha) &=& 1+ \delta m_\tau^{(0)}(T,\alpha) n_B + { \mathcal{O}}(n_B^2)\,.
\end{eqnarray}

To illustrate the quasiparticle approach in the single-nucleon case, we give the Skyrme I parametrization by Vautherin and Brink \cite{Vau72,LS} which represents an analytical expression for the quasiparticle shift of the single nucleon states, 
\begin{equation}
\Delta E^{\rm SE}_{n}(0)= \frac{t_0}{2} \left(1-x_0\right)n_n+\frac{t_0}{2} \left(2+x_0\right)n_p+\frac{t_3}{4} \left(2 n_n n_p+n_p^2\right)+\left( \frac{t_1}{8}+ \frac{3t_2}{8}\right)\tau_n+ \left(\frac{t_1}{4}+ \frac{t_2}{4}\right)\tau_p\,,
\end{equation}
and the effective mass
\begin{equation}
\frac{m_n^*}{m_n} = \left\{ 1 +\left[ \left( \frac{t_1}{4}+ \frac{3 t_2}{4}\right)n_n+ \left(\frac{t_1}{2}+ \frac{t_2}{2}\right)n_p  \right] \frac{m_n}{\hbar^2}   \right\}^{-1}\,,
\end{equation}
with the parameter values $t_0=-1057.3$ MeV fm$^3$, $t_1=235.9$ MeV fm$^5$,  $t_2=-100$ MeV fm$^5$,  $t_3= 14463.5$ MeV fm$^6$, $x_0 =0.56$. $\tau_n$ denotes the neutron kinetic energy per particle. Expressions for the protons are found by interchanging $n$ and $p$. 
For symmetric matter we find the shift $ \delta E_\tau(T,0)=-\frac{3}{4}$ 1057.3 MeV fm$^3$.
The temperature dependence is only weak.

Recently, the EoS of asymmetric nuclear matter has been investigated in the low-density region below the nuclear saturation density \cite{mar07}. Microscopic calculations based on the Dirac-Brueckner Hartree Fock approach with realistic nucleon- nucleon potentials are used to adjust a low-density energy functional. This functional is constructed on a density expansion of the relativistic mean-field theory.
An improved version of the relativistic mean field approach to reproduce properties of nuclei in wide range of $A$ and $Z$ was worked out by Typel \cite{Typel}. 
The relativistic quasiparticle energy (\ref{RMFquasi}) reads in the non-relativistic limit $\Delta E^{\rm SE}_{\tau}(0)=-S(n_n,n_p,T)+V_\tau(n_n,n_p,T)$ and $m_\tau^*/m_\tau = 1-S(n_n,n_p,T)/(m_\tau c^2)$.
For the low-density limit, expansions of the scalar and vector potentials with respect to the neutron and proton densities can be given.

\subsection{Quasideuteron}

In the deuteron case, we introduce Jacobi coordinates such that $q_1=(p_2-p_1)/2=p_{\rm rel}, q_2=p_1+p_2 =P$ or $p_1=-q_1+q_2/2, \, p_2=q_1+q_2/2$.
The kinetic energy is $KE = \frac{\hbar^2}{2m} (2q_1^2 + \frac{1}{2} q_2^2)$, the interaction is parametrized by the separable Gaussian (\ref{Gausspot})
\begin{equation}
V(q_1,q_2,q_1',q_2') = \lambda e^{-\frac{q_1^2}{\gamma^2}} e^{-\frac{{q_1'}^2}{\gamma^2}} \delta_{q_2,q_2'} \,.
\end{equation} 
The deuteron wave function (\ref{varphi}) results as $\varphi_2(q_1) \propto e^{-2 q_1^2/a^2}/(q_1^2/b^2+1)$.
The value of the rms radius follows as
\begin{equation}
{\rm rms}^2 = \frac{1}{2}  \langle[(r_1-R)^2+(r_2-R)^2]\rangle =
\frac{1}{4}  \int d^3 q_1 \left[ \frac{ \partial}{\partial q_1} \varphi_d(q_1,q_2)\right]^2 \,.
 \end{equation}

As in the case of single nucleons, the quasiparticle shifts can be expanded as power series of the densities,
\begin{equation}
\Delta E_d(P;T,n_B,\alpha) = \delta E_d(P;T,\alpha) n_B + { \mathcal{O}}(n_B^2)
\end{equation}
The first order term $  \delta E_d $ consists of the self-energy contribution and the Pauli-blocking contribution, see Sec. \ref{subsub3.2.2}, in particular 
$\delta E_d^{\rm rigid \, shift}= \delta E_p(0;T,\alpha)  + \delta E_n(0;T,\alpha)$.
Furthermore we have
\begin{equation}
\frac{m^*_d}{m_d}(T,n_B,\alpha) =1+ \delta m_d(T,\alpha) n_B + { \mathcal{O}}(n_B^2)\,.
\end{equation}

Values for $\delta E_d^{\rm Pauli}(0;T)$ are given in Tab. \ref{Tab.3}. For $T$=10 MeV we have $\delta m_d(10,0)$=21.3 fm$^3$, whereas for $T$=4 MeV  the value $\delta m_d(4,0)$=87.1 fm$^3$ results. Due to the Pauli blocking both quantities are strongly temperature dependent.
At zero temperature and low densities, we find for the Gaussian interaction
\begin{equation}
\delta E_d^{\rm Pauli}(P;0) =- \frac{1}{2}  \lambda_d  \psi_d(P/2)  e^{-\frac{P^2}{4 \gamma_d^2}} \frac{ \int d^3q_1  e^{-q_1^2/\gamma_d^2} \psi_d(q_1) }{ \int d^3q_1 | \psi_d(q_1)|^2}\,.
\end{equation}

\subsection{Quasitriton/helion}

Next we consider $A=3\,\, (t,h)$. Jacobi coordinates are $q_1=\frac{1}{2}(p_2-p_1), q_2=\frac{2}{3}(-\frac{1}{2}p_1-\frac{1}{2}p_2 +p_3), q_3=p_1+p_2+p_3$ or $p_1=-q_1-\frac{1}{2}q_2+\frac{1}{3} q_3, \,p_2=q_1-\frac{1}{2}q_2+\frac{1}{3} q_3, \, p_3=q_2+\frac{1}{3} q_3$.
The kinetic energy is $KE = \frac{\hbar^2}{2m} (2q_1^2 + \frac{3}{2} q_2^2 + \frac{1}{3} q_3^2)$.

We start from a Gaussian pair interaction (\ref{Gausspot})
which gives in Jacobian coordinates the three-nucleon interaction
\begin{eqnarray}
&& V^{\rm pair}_3(q_1,q_2,q_3,q_1',q_2',q_3') = \lambda \delta_{q_3,q_3'} \left\{  \delta_{q_2,q_2'} e^{-\frac{q_1^2}{\gamma^2}} e^{-\frac{{q_1'}^2}{\gamma^2}} 
\right.\nonumber \\ && \left.
+ \delta_{q_1',q_1+\frac{q_2'-q_2}{2}} e^{-\frac{(q_1+\frac{3}{2} q_2)^2}{4 \gamma^2}} e^{-\frac{(q_1-\frac{1}{2} q_2+2 q_2')^2}{4 \gamma^2}}
+ \delta_{q_1',q_1-\frac{q_2'-q_2}{2}} e^{-\frac{(q_1-\frac{3}{2} q_2)^2}{4 \gamma^2}} e^{-\frac{(q_1+\frac{1}{2} q_2-2 q_2')^2}{4 \gamma^2}} \right\}\,.
\end{eqnarray}

The Gaussian variational ansatz (\ref{varGauss}) reads after introducing Jacobians (indices in  $B,a,b$ are omitted)
\begin{equation}
 \varphi^{\rm Gauss}_3(q_1,q_2,q_3) \propto e^{-\frac{2 q_1^2}{B^2}}  e^{-\frac{3 q_2^2}{2 B^2}} \delta_{q_3,P}\,.
\end{equation}

The Jastrow variational ansatz (\ref{varphi}) motivated by the solution of the two-particle problem,
reads after introduction of the reduced Jacobian coordinates $\vec x_i = \vec q_i/b$ and choosing the coordinates as $\vec x_1=x_1\{(1-z^2)^{1/2},0,z\}, \, \vec x_2=x_2\{0,0,1\}$
\begin{eqnarray}
\varphi^{\rm Jastrow}_3(x_1,x_2,z) \propto \frac{e^{-\frac{3}{2}\frac{b^2}{a^2}x_1^2-\frac{9}{8}\frac{b^2}{a^2}x_2^2} } {(x_1^2+1) (\frac{1}{4}x_1^2+\frac{9}{16} x_2^2+\frac{3}{4} x_1 x_2 z+1)(\frac{1}{4}x_1^2+\frac{9}{16} x_2^2-\frac{3}{4} x_1 x_2 z+1)}
\end{eqnarray}

The kinetic energy follows as 
\begin{equation}
KE_3=\frac{\hbar^2}{m} \frac{b^2}{N_3} \int d x_2\,x_2^2  \int d x_1\,x_1^2 \left(x_1^2+ \frac{3}{4}x_2^2\right) \int_{-1}^1 dz \varphi_3^2(x_1,x_2,z)
\end{equation}
with the norm
\begin{equation}
N_3= \int d x_2\,x_2^2  \int d x_1\,x_1^2  \int_{-1}^1 dz \phi_3^2(x_1,x_2,z)\,.
\end{equation}
For the potential energy we obtain
\begin{equation}
PE_3=3 \lambda_3 \frac{b^3}{4 \pi^2 N_3} \int d x_2\,x_2^2  \left[ \int d x_1\,x_1^2 e^{-x_1^2 \frac{b^2}{\gamma^2}} \int_{-1}^1 dz \varphi_3(x_1,x_2,z)\right]^2\,.
\end{equation}

The nucleonic point rms radius follows as
\begin{equation}
 {\rm rms}_3^2= \frac{1}{b^2 N_3} \int d x_2\,x_2^2  \int d x_1\,x_1^2  \int_{-1}^1 dz \left[ \frac{1}{6} \left( \frac{ \partial \varphi_3}{\partial \vec x_1}\right)^2 
+ \frac{2}{9} \left( \frac{ \partial \varphi_3}{\partial \vec x_2}\right)^2 \right]\,,
\end{equation}
in particular rms$^2_3 = 2/B^2$ for the Gaussian ansatz (\ref{varGauss}).

At finite temperature, the Pauli blocking contribution to the quasiparticle shift is given by
\begin{eqnarray}
&&\delta E^{\rm Pauli,J}_3(P) =\frac{3 \lambda_3 b^3}{8 \pi^2 N_3}  \left(\frac{2 \pi \hbar^2}{m T}\right)^{3/2}  \int d x_2\, x_2^2 \\ && \times \int d x_1\, x_1^2 \int dz_1 \varphi_3(x_1,x_2,z_1)  e^{-x_1^2 \frac{b^2}{ \gamma^2}} e^{- \frac{\hbar^2 b^2}{2 m T} (\vec x_1+ \frac{1}{2}\vec x_2 -\frac{1}{3} \vec x_3)^2} \int d x_5\, x_5^2 \int dz_5 \varphi_3(x_5,x_2,z_5)  e^{-x_5^2 \frac{b^2}{ \gamma^2}}\,.\nonumber
\end{eqnarray}
At zero temperature where $p_1 \approx 0$ ($\vec x_1=\vec x_3/3-\vec x_2/2;\,\,\vec x_3=0 $), the shift is given by
\begin{equation}
\delta E^{\rm Pauli,J}_3(0)=\frac{3 \lambda_3}{2 N_3} \int d x_2\, x_2^2 \frac{e^{-x_2^2 (\frac{3b^2}{2 a^2} +\frac{b^2}{4 \gamma^2})}}{(x_2^2+1)(\frac{x_2^2}{4}+1)^2} \int d x_5\, x_5^2 \int dz_5 \varphi_3(x_5,x_2,z_5)  e^{-x_5^2 \frac{b^2}{ \gamma^2}}\,.
\end{equation}

\subsection{The $\alpha$-quasiparticle}

To solve the four-nucleon Schr\"odinger equation 
in the zero-density limit, we separate the center-of-mass motion from the internal motion introducing Jacobian coordinates,
$q_1=-\frac{1}{2}p_1+\frac{1}{2}p_2, q_2=-\frac{1}{3}p_1-\frac{1}{3}p_2+\frac{2}{3} p_3,  q_3=-\frac{1}{4}p_1-\frac{1}{4}p_2-\frac{1}{4}p_3+\frac{3}{4} p_4, q_4=p_1+p_2+p_3+p_4$. The inverse transformation is $p_1=\frac{1}{4} q_4-\frac{1}{3}q_3-\frac{1}{2}q_2-q_1, \,p_2=\frac{1}{4} q_4-\frac{1}{3}q_3-\frac{1}{2}q_2+q_1,p_3=\frac{1}{4} q_4-\frac{1}{3}q_3+q_2, \, p_4=\frac{1}{4} q_4+ q_3$.

The Schr\"odinger equation separates with $\varphi_{\alpha,P}(1,2,3,4)=\varphi_4(\vec q_1,\vec q_2,\vec q_3) \delta_{\vec P,\vec q_4}$. The kinetic energy is $KE_4 = \frac{\hbar^2}{2m} (2q_1^2 + \frac{3}{2} q_2^2 + \frac{4}{3} q_3^2 + \frac{1}{4} P^2)$. The potential energy follows as
\begin{eqnarray}
&&V^{\rm pair}_4(q_1,q_2,q_3,q_4,q_1',q_2',q_3',q_4') = \lambda_\alpha \delta_{q_4,q_4'} \left\{  \delta_{q_3,q_3'} \delta_{q_2,q_2'} e^{-\frac{q_1^2}{ \gamma_\alpha^2}} e^{-\frac{{q_1'}^2}{ \gamma_\alpha^2}} 
\right.\nonumber \\ && \left.
+  \delta_{q_3,q_3'}\delta_{q_1',q_1+\frac{1}{2}(q_2'-q_2)} e^{-\frac{(q_1+\frac{3}{2} q_2)^2}{4  \gamma_\alpha^2}} e^{-\frac{(q_1-\frac{1}{2} q_2+2 q_2')^2}{4  \gamma_\alpha^2}}
+  \delta_{q_3,q_3'}\delta_{q_1',q_1-\frac{1}{2}(q_2'-q_2)} e^{-\frac{(q_1-\frac{3}{2} q_2)^2}{4  \gamma_\alpha^2}} e^{-\frac{(q_1+\frac{1}{2} q_2-2 q_2')^2}{4  \gamma_\alpha^2}}
\right.\nonumber \\ && \left.
+  \delta_{q_1,q_1'}\delta_{q_2',q_2+\frac{2}{3}(q_3-q_3')} e^{-\frac{(-q_2+\frac{4}{3} q_3)^2}{4  \gamma_\alpha^2}} e^{-\frac{(-3 q_2-2 q_3+6 q_3')^2}{36  \gamma_\alpha^2}}
\right.\nonumber \\ && \left.
+  \delta_{q_1',q_1-\frac{1}{2}(q_3-q_3')}\delta_{q_2',q_2-\frac{1}{3}(q_3-q_3')} e^{-\frac{(q_1+\frac{1}{2} q_2+\frac{4}{3} q_3)^2}{4  \gamma_\alpha^2}} e^{-\frac{(6 q_1 +3 q_2 -4 q_3+12 q_3')^2}{144  \gamma_\alpha^2}}
\right.\nonumber \\ && \left.
+  \delta_{q_1',q_1+\frac{1}{2}(q_3-q_3')}\delta_{q_2',q_2-\frac{1}{3}(q_3-q_3')} e^{-\frac{(-q_1+\frac{1}{2} q_2+\frac{4}{3} q_3)^2}{4  \gamma_\alpha^2}} e^{-\frac{(-6 q_1 +3 q_2 -4 q_3+12 q_3')^2}{144  \gamma_\alpha^2}}
 \right\}\,.
\end{eqnarray}

To solve the internal motion we use a variational ansatz for the wave function. The Gaussian variational ansatz (\ref{varGauss}) reads after introducing Jacobians
\begin{equation}
 \varphi^{\rm Gauss}_\alpha(q_1,q_2,q_3,q_4) \propto e^{-\frac{2 q_1^2}{B^2}}  e^{-\frac{3 q_2^2}{2 B^2}} e^{-\frac{4 q_3^2}{3 B^2}} \delta_{q_4,P}\,.
\end{equation}

The Jastrow variational ansatz (\ref{varphi})
\begin{equation}
 \varphi^{\rm Jastrow}_{\alpha,P}(\vec p_1,\vec p_2,\vec p_3,\vec p_4) \propto \frac{e^{-\frac{1}{4 a^2}(p_2-p_1)^2}}{\frac{(p_2-p_1)^2}{4 b^2}+1} \dots 
\frac{e^{-\frac{1}{4 a^2}(p_4-p_3)^2}}{\frac{(p_4-p_3)^2}{4 b^2}+1}\,\,\delta_{\vec p_1+\vec p_2+\vec p_3+\vec p_4,\vec P} \,.
\end{equation}
reads after introduction of the reduced Jacobian momenta  $\vec x_i=\vec q_i/b$, with $\vec x_1 = x_1 \{\sqrt{1-z_1^2} \cos (\phi_1),\sqrt{1-z_1^2} \sin (\phi_1), z_1\}; \vec x_2 = x_2 \{0,0,1\}; \vec x_3 = x_3 \{\sqrt{1-z_3^2},0, z_3\}$,
\begin{eqnarray}
&&\varphi_4(x_1,z_1,\phi_1,x_2,x_3,z_3)=\frac{e^{-\frac{b^2}{a^2}(2 x_1^2+\frac{3}{2}x_2^2+\frac{4}{3}x_3^2)}}{(x_1^2+1) [(\frac{1}{4}x_1^2+\frac{9}{16}x_2^2+1)^2-\frac{9}{16} x_1^2 x_2^2 z_1^2] (\frac{1}{4}x_2^2+\frac{4}{9}x_3^2-\frac{2}{3} x_2 x_3 z_3+1)}\nonumber \\&&\frac{1}{[(\frac{1}{4}x_1^2+\frac{1}{16}x_2^2 +\frac{4}{9}x_3^2+\frac{1}{3} x_2 x_3 z_3+1)^2-(\frac{1}{4}x_1 x_2 z_1+\frac{2}{3} x_1 x_3 \{z_1 z_3+ \sqrt{1-z_1^2}  \sqrt{1-z_3^2} \cos \phi_1\})^2]}\,.\nonumber \\&&{}
\end{eqnarray}
We evaluate the norm as
\begin{equation}
N_4 = \int_0^\infty d x_1 x_1^2  \int_{-1}^1 \frac{dz_1}{2}\int_0^{2 \pi} \frac{d \phi_1}{2 \pi} \int_0^\infty d x_2 x_2^2 \int_0^\infty d x_3 x_3^2 \int_{-1}^1 \frac{dz_3}{2} \varphi_4^2\,.
\end{equation}
The kinetic energy of the internal motion is calculated from
\begin{eqnarray}
{\rm KE}_4 & = & \frac{\hbar^2 b^2}{m \,\,N_4}  \int_0^\infty d x_1 x_1^2  \int_{-1}^1 \frac{dz_1}{2}\int_0^{2 \pi} \frac{d \phi_1}{2 \pi} \int_0^\infty d x_2 x_2^2 \int_0^\infty d x_3 x_3^2 \int_{-1}^1 \frac{dz_3}{2}\nonumber \\&& \times \left(x_1^2 +\frac{3}{4} x_2^2+\frac{2}{3} x_3^2\right) \varphi_4^2
\end{eqnarray}
The potential energy is (all six terms give the same contribution; we take the first one as representative)
\begin{equation}
{\rm PE}_4 = \frac{6 \lambda b^3}{2 \pi^2  N_4} \int_0^\infty d x_2 x_2^2 \int_0^\infty d x_3 x_3^2\int_{-1}^1 \frac{dz_3}{2}  \left( \int_0^\infty d x_1 x_1^2  \int_{-1}^1 \frac{dz_1}{2}  \int_0^{2 \pi} \frac{d \phi_1}{2 \pi} e^{-\frac{b^2}{\gamma^2}x_1^2} \varphi_4 \right)^2\,.
\end{equation}
The integrals over $\phi_1$ can be performed analytically.

The nucleonic point rms radius follows as
\begin{eqnarray}
 {\rm rms}_4^2&=& \frac{1}{b^2 N_4} \int_0^\infty d x_1 x_1^2  \int_{-1}^1 \frac{dz_1}{2}\int_0^{2 \pi} \frac{d \phi_1}{2 \pi} 
\int_0^\infty d x_2 x_2^2 \int_0^\infty d x_3 x_3^2 \int_{-1}^1 \frac{dz_3}{2}\nonumber \\&& \left[ \frac{1}{8} \left( \frac{ \partial \varphi_4}{\partial \vec x_1}\right)^2 
+ \frac{1}{6} \left( \frac{ \partial \varphi_4}{\partial \vec x_2}\right)^2+ \frac{3}{16} \left( \frac{ \partial \varphi_4}{\partial \vec x_3}\right)^2 \right]\,,
\end{eqnarray}
in particular rms$^2 = 9/4 B^2$ for the Gaussian ansatz (\ref{varGauss}).

Calculating the rms radius for the Jastrow ansatz, the three terms give the same contribution so that we take the last one as representative
\begin{eqnarray}
&&{\rm rms}_4^2 = \frac{9}{16 \,\,b^2\,N_4}   \int_0^\infty d x_1 x_1^2  \int_{-1}^1 \frac{dz_1}{2}\int_0^{2 \pi} \frac{d \phi_1}{2 \pi} \int_0^\infty d x_2 x_2^2 \int_0^\infty d x_3 x_3^2 \int_{-1}^1 \frac{dz_3}{2}  \varphi_4^2 \frac{1}{f_1^2 (f_2^2-f_3^2)^2}
\nonumber\\&& \times \left\{ x_3^2 \left[\frac{8 b^2}{3 a^2} f_1 (f_2^2-f_3^2)+\frac{32}{9} (f_2^2-f_3^2)+ \frac{64}{9} f_1 f_2 \right]^2 +x_2^2 \left[-\frac{8}{3 }  (f_2^2-f_3^2)+\frac{8}{3}  f_1 f_2 \right]^2  +x_1^2 \left[-\frac{16}{3 }   f_1 f_3 \right]^2 
\right.
\nonumber\\&&  \left. + 2 x_2 x_3 z_3 \left[\frac{8 b^2}{3 a^2} f_1 (f_2^2-f_3^2)+\frac{32}{9} (f_2^2-f_3^2)+ \frac{64}{9} f_1 f_2\right] \left[-\frac{8}{3 }  (f_2^2-f_3^2)+\frac{8}{3}  f_1 f_2 \right] \right.
\nonumber\\&&  \left. +2 x_1 x_3 (z_1 z_3+ \sqrt{1-z_1^2}  \sqrt{1-z_3^2} \cos \phi_1) \left[\frac{8}{3 \beta^2} f_1 (f_2^2-f_3^2)+\frac{32}{9} (f_2^2-f_3^2)+ \frac{64}{9} f_1 f_2\right] \left[-\frac{16}{3 }  f_1 f_3 \right]  \right.
\nonumber\\&&  \left. +2 x_1 x_2 z_1 \left[-\frac{8}{3 }  (f_2^2-f_3^2)+\frac{8}{3}  f_1 f_2 \right] \left[-\frac{16}{3 }  f_1 f_3 \right]
\right\}\,,
\end{eqnarray}
where we used the abbreviations $f_1= (\frac{1}{4}x_2^2+\frac{4}{9}x_3^2-\frac{2}{3} x_2 x_3 z_3+1), f_2=(\frac{1}{4}x_1^2+\frac{1}{16}x_2^2 +\frac{4}{9}x_3^2+\frac{1}{3} x_2 x_3 z_3+1), f_3=(\frac{1}{4}x_1 x_2 z_1+\frac{2}{3} x_1^2 x_3^2 \{z_1 z_3+ \sqrt{1-z_1^2}  \sqrt{1-z_3^2} \cos \phi_1\})$.

At finite densities, we take into account self-energies and Pauli blocking due to the medium. 
We obtain 12 terms which differ only by the isospin ($p,n$) dependence so that
\begin{equation}
\Delta E^{\rm Pauli}_\alpha(P)= -6 \sum_{1234,1'2'}   \varphi^*_{\alpha,P}(1,2,3,4)[f_p(1)+f_n(1)]V(1,2;1',2')  \varphi_{\alpha,P}(1',2',3,4)\,.
\end{equation}
First we consider $P = 0$.
At $T=0$, the Fermi distribution function is replaced by the step function. In the low-density limit where $\vec p_1=0 $ or $\vec x_1=-\frac{1}{2} \vec x_2-\frac{1}{3} \vec x_3$, we have
\begin{eqnarray}
&&  \delta E^{\rm Pauli,J}_4(0;0) =- \frac{6 \lambda}{2 \,\, N_4}\int_{-1}^1 \frac{dz_3}{2} \int_0^\infty d x_3 x_3^2 \int_0^\infty d x_2 x_2^2 
\nonumber\\ && \times
\frac{e^{-\frac{b^2}{a^2}(2 x_2^2+\frac{14}{9}x_3^2+\frac{2}{3}x_2 x_3 z_3)} e^{-\frac{b^2}{\gamma^2}(\frac{1}{4}x_2^2+\frac{1}{9}x_3^2+\frac{1}{3}x_2 x_3 z_3)} }{(x_2^2+\frac{4}{9}x_3^2+\frac{4}{3}x_2 x_3 z_3+1) [(\frac{5}{2}x_2^2+\frac{1}{9}x_3^2+\frac{1}{3}x_2 x_3 z_3+1)^2-9 (\frac{1}{2}x_2^2 +\frac{1}{3}x_2 x_3 z_3)^2]}  
\nonumber\\ && \times \frac{1}{ (x_2^2+\frac{16}{9}x_3^2-\frac{8}{3} x_2 x_3 z_3+1)[(\frac{1}{2}x_2^2+\frac{17}{9}x_3^2+\frac{5}{3}x_2 x_3 z_3+1)^2- (\frac{1}{2}x_2^2 +\frac{8}{9}x_3^2 +\frac{5}{3}x_2 x_3 z_3)^2]}
 \nonumber\\ && \times \left\{ \int_0^\infty d x_5 x_5^2  \int_{-1}^1 \frac{dz_5}{2}   \int_0^{2 \pi} \frac{d \phi_5}{2 \pi} e^{-\frac{b^2}{\gamma^2}x_5^2} \varphi_4(x_5,x_2,x_3,z_5,z_3,\phi_5) \right\}\,.
\end{eqnarray}
The integral over $\phi_5$ can be performed analytically.

For arbitrary temperatures we find in the low-density limit, where we have $f_\tau(p) = \frac{1}{2} n_\tau (2 \pi \hbar^2/m T)^{3/2} \exp[ -\frac{\hbar^2}{2 m T} p^2]$, the expression
\begin{eqnarray}
&&  \delta E^{\rm Pauli,J}_4(0;T) =- \frac{3 \lambda b^3}{2 \pi^2 N_4} \left(\frac{2 \pi \hbar^2}{m\,T}\right)^{3/2}\int_{-1}^1 \frac{dz_3}{2} \int_0^\infty d x_3 x_3^2 \int_0^\infty d x_2 x_2^2 \\ && \times
\left\{ \int_{-1}^1 \frac{dz_1}{2}  \int_0^\infty d x_1 x_1^2  \int_0^{2 \pi} \frac{d \phi_5}{2 \pi} 
 \right. \nonumber\\ && \left. \times   e^{-\frac{b^2}{\gamma^2}x_1^2} e^{-\frac{\hbar^2 b^2}{2 m T} (x_1^2+\frac{1}{4} x_2^2+\frac{1}{9} x_3^2+x_1 x_2 z_1+\frac{1}{3} x_2 x_3 z_3+\frac{2}{3} x_1 x_3( z_1 z_3+ \sqrt{1-z_1^2}  \sqrt{1-z_3^2} \cos (\phi_1))} \right. \nonumber\\ && \left. \times  \varphi_4(x_1,x_2,x_3,z_1,z_3,\phi_1) \right\} \left\{ \int_0^\infty d x_5 x_5^2  \int_{-1}^1 \frac{dz_5}{2}   \int_0^{2 \pi} \frac{d \phi_5}{2 \pi} e^{-\frac{b^2}{\gamma^2}x_5^2} \varphi_4(x_5,x_2,x_3,z_5,z_3,\phi_5)\right\} \,. \nonumber
\end{eqnarray}

For finite $P$ we have to introduce in the Fermi distribution  $p_1/b=\frac{1}{4} \vec P/b-\frac{1}{3} \vec x_3-\frac{1}{2} \vec x_2- \vec x_1$. Expansion for small $P$ gives the Pauli blocking contribution to the effective mass.

\end{document}